\pdfoutput=1
\RequirePackage[T1]{fontenc}
\RequirePackage{fix-cm}
\RequirePackage{lmodern}

\documentclass[11pt,a4paper,english]{amsart}
\usepackage{babel}
\usepackage[latin1]{inputenc}
\usepackage{amsmath}
\usepackage{amsthm}
\usepackage{amsfonts}
\usepackage{indentfirst}
\usepackage{graphicx}
\usepackage{amssymb}
\usepackage[mathscr]{eucal}
\usepackage{tikz}
\usepackage{caption}

\usepackage{enumerate}
\usepackage{amsthm}

\usepackage[colorlinks=true,citecolor=black,linkcolor=black,urlcolor=blue]{hyperref}
\usepackage{tikz,graphics,color,fullpage,float,epsf,caption,subcaption}
\usepackage{amscd}
\newtheorem{teo}{Theorem}

\newtheorem{pro}[teo]{Proposition}
\newtheorem{lem}[teo]{Lemma}

\theoremstyle{definition}
\newtheorem{rem}[teo]{Remark}
\newtheorem{de}[teo]{Definition}
\newtheorem{exa}{Example}

\newcommand{\ket}[1]{|#1 \rangle}
\newcommand{\bra}[1]{\langle #1 |}

\title[Quantum $(r,\delta)$-Locally Recoverable Codes]{Quantum $(r,\delta)$-Locally Recoverable Codes}
\author[C. Galindo, F. Hernando, H. Mart\'{\i}n-Cruz and R. Matsumoto]{Carlos Galindo, Fernando Hernando, Helena Mart\'{\i}n-Cruz and Ryutaroh Matsumoto}
\curraddr{\texttt{Carlos Galindo and Fernando Hernando:} Instituto
Universitario de Matem\'aticas y Aplicaciones de Castell\'on and
Departamento de Matem\'aticas, Universitat Jaume I, Campus de Riu
Sec. 12071 Castell\'{o} (Spain)\\ \texttt{Helena Mart\'{\i}n-Cruz:} Departamento de Matem\'aticas, Universidad de Ja\'en, Campus Las Lagunillas. 23071 Ja\'en (Spain)\\ 
\texttt{Ryutaroh Matsumoto:} Department of Information and Communications Engineering, Institute of Science Tokyo, Japan.}

\email{{\rm Galindo: galindo@uji.es; {\rm Hernando:} carrillf@uji.es; {\rm Mart\'{\i}n-Cruz:} hmartin@ujaen.es; {\rm Matsumoto:} ryutaroh@ict.e.titech.ac.jp}}

\urladdr{{\rm Galindo: 0000-0002-3908-4462; {\rm Hernando:} 0000-0002-9758-2152; {\rm Mart\'{\i}n-Cruz:} 0000-0002-6379-6902; {\rm Matsumoto:} 0000-0002-5085-8879}\color{black}}

\thanks{The first three authors were partially funded by MCICIU/AEI/10.13039/501100011033 and by ``ERDF, UE'' (grant PID2022-138906NB-C22), and by University Jaume I (grant GACUJIMA/2024/03). The third author was also partially funded by University Jaume I (grant PREDOC/2020/39). The fourth author was partially supported by ``Japan Society of the Promotion of the Science'' (grant 23K10980). }

\begin{document}

\begin{abstract}
Classical $(r,\delta)$-locally recoverable codes are designed for avoiding loss of information in large scale distributed and cloud storage systems. We introduce the quantum counterpart of those codes by defining quantum $(r,\delta)$-locally recoverable codes which are quantum error-correcting codes capable of correcting $\delta -1$ qudit erasures from sets of at most $r+ \delta -1$ qudits.

We give a necessary and sufficient condition for a quantum stabilizer code $Q(C)$ to be $(r,\delta)$-locally recoverable. Our condition depends only on the puncturing and shortening at suitable sets of both the symplectic self-orthogonal code $C$ used for constructing $Q(C)$ and its symplectic dual $C^{\perp_s}$. When $Q(C)$ comes from a Hermitian or Euclidean dual-containing code, and under an extra condition, we show that there is an equivalence between the classical and quantum concepts of $(r,\delta)$-local recoverability. A Singleton-like bound is stated in this case and examples attaining the bound are given.
\end{abstract}

\maketitle

\section{Introduction}\label{se:uno}
Literature contains various sources giving some evidence of quantum supremacy  \cite{Aru,BallP,Zhong,LiuY}, which makes
that quantum information processing have so much potential for multiple practical applications. When using quantum devices, it is important to protect quantum information from decoherence and other quantum noise. {\it Quantum error-correcting codes} (QECCs) are designed to perform this task since quantum correction works  \cite{ShorS,95kkk} despite the fact that one cannot clone quantum information \cite{Dieks,Woot}.

Only binary codes were considered in the seminal contributions to the study of QECCs \cite{20kkk,Gottesman,Calder2,Calderbank}. At a later stage, $q$-ary QECCs ($q$ being a power of a prime number $p$) have been extensively studied and constructed with a wide variety of techniques, see for instance \cite{BE, AK, Ketkar, Aly, XingC, Lag2, gahe, Anderson} among many other papers on the subject.

A QECC of length $n$ is simply a $K$-dimensional linear subspace of the complex Hilbert space $\mathbb{C}^q \otimes \cdots \otimes \mathbb{C}^q = \mathbb{C}^{q^n}$. {\it Quantum stabilizer codes} constitute one of the best-known class of QECCs. They are particularly interesting because of their relation with classical codes. The goodness of (quantum) stabilizer codes is measured by their  parameters $((n,K,d))_q$, where $d$ means that the code  can detect all errors in the error group of weight less than $d$ but not some of weight $d$.
Let $\mathbb{F}_q$ be the finite field of cardinality $q$, an $((n,K,d))_q$ stabilizer code exists if and only if there is an additive classical code $C \subseteq \mathbb{F}_q^{2n}$ which is self-orthogonal with respect to the trace-symplectic form. In addition, when $K>1$, $d$ coincides with the symplectic weight of $C^{\perp_{\mathrm{trs}}} \setminus C$, $C^{\perp_{\mathrm{trs}}}$ being the dual of $C$ with respect to the above form. This relation between quantum and classical error-correcting codes simplifies the obtention and study of stabilizer codes. If one uses the easier to handle Hermitian, or Euclidean, inner product on $\mathbb{F}_{q^2}^{n}$, or on $\mathbb{F}_{q}^{n}$, quantum stabilizer  codes (and their parameters) can be gotten from self-orthogonal classical codes with respect to the mentioned products (and their parameters). In the Euclidean case, one obtains the so-called CSS (quantum) codes, see Item (2) in the forthcoming Theorem \ref{resto}.


Many big companies store information in large scale distributed and cloud storage systems and, for them, the loss of the data in a node is a major issue. The so-called repair problem proposes to recover the lost data from the information deposited in the remaining nodes. To do it, in \cite{GHSY2012} the authors suggest to use classical error-correcting codes (ECCs). Thus, we say that an ECC has $r$-locality when one can recover any coordinate of a word in $C$ by accessing at most $r$ other symbols of the word. These codes are named {\it $r$-locally recoverable codes} ($r$-LRCs) or LRCs with locality $r$. Some recent references on them are \cite{TPD2016,BTV2017, LMC2018,Mi2018,LXY2019,J2019,LMT2020,SVV2021, edgar}. A Singleton-like bound for $r$-LRCs was proposed in \cite{GHSY2012}, see (\ref{Singleton1}). Optimal $r$-LRCs are those reaching that bound. Many of the above cited papers contain families of optimal LRCs.

Locally recoverable codes with locality $r$ are only designed for the case when a node fails, but simultaneous multiple device failures are not infrequent. To address this issue, {\it $(r,\delta)$-locally recoverable codes} ($(r,\delta)$-LRCs), also named LRCs with locality $(r,\delta)$, were introduced in \cite{PKLK2012}. Fixed a coordinate $c_i$ of a codeword, these codes are capable of correcting $\delta-1$ erasures in a set containing $c_i$ plus, at most, $r + \delta -2$ other coordinates. These codes also admit a Singleton-like bound (\ref{Singleform}) and when reached they are called optimal $(r,\delta)$-LRCs. A considerable effort has been made recently to get optimal $(r,\delta)$-LRCs \cite{SDYL2014, CXHF2018,LMX2019,LXY2019,SZW2019,J2019,CFXF2019,Z2020,FF2020,QZF2021,KWG2021, Cai, GFMC}. 

The good prospects offered by quantum computing and the growing amount of information stored by data centers make it reasonable to introduce quantum locally recoverable codes. The large budget of companies such as Microsoft and Meta, could make it feasible to use  quantum computers to store information, which would require local recovery. Thus, quantum locally recoverable codes (quantum LRCs) would be the quantum counterpart of classical LRCs. The first foray into this field occurred in \cite{Golowich} (see also \cite{Golowich2}) where the authors introduce {\it quantum $r$-locally recoverable  codes} (quantum $r$-LRCs). Similarly to the classical case, these codes are QECCs that correct an erasure corresponding to an input qudit. A condition for a quantum stabilizer code to be a quantum $r$-LRC and its particularization to local recoverability of CSS codes is provided in  \cite{Golowich}.
Despite the recent introduction of quantum $r$-LRCs, one can find contributions in this last line  \cite{LuoG, Sharma}. In addition, \cite{Golowich}, among other results, contains a Singleton-like bound, which admits a stronger version under a specific behavior of the recovery sets \cite[Theorem 36]{Golowich}. Next, we reproduce that last bound for an $[[n,k,d]]_q$ quantum $r$-LRC:
\begin{equation}
\label{SingletonRQ}
k \leq \left( 1 -
 \frac{2}{r+1} \right) n - 2 \left( d-1 - \left\lceil \frac{d-1}{r} \right\rceil \right).
\end{equation}

To complete the quantum counterpart of classical LRCs, it remained to introduce quantum $(r,\delta)$-LRCs, task which is addressed in this paper. Given a QECC, $Q \subseteq \mathbb{C}^{q^n}$, and two sets of indices $I$ and $J$ such that $\emptyset \subsetneq I \subsetneq J \subseteq\{1, \ldots, n\}$, roughly speaking, we say that $Q$ is $(I,J)$-locally recoverable whenever $Q$ corrects erasures at (the positions in) $I$ from the qudits of $Q$ at $J$. See the more accurate Definition \ref{I-J}, where we introduce the concept of $(I,J)$-local recoverability with the aim of reverting the action of an arbitrary error channel (in fact, a completely depolarizing channel) on qudits at $I$ by using qudits at $J$. Then, a
QECC  $Q \subseteq \mathbb{C}^{q^n}$ is a {\it quantum $(r,\delta)$-locally recoverable code} (quantum $(r,\delta)$-LRC) if for each index $i \in \{1, \ldots, n\}$, there exists a set $J \subseteq \{1, \ldots, n\}$ containing $i$ with $\mathrm{card}\,(J) \leq r + \delta-1$ such that, for all subset $I \subseteq J$ of cardinality $\delta -1$, $Q$ is $(I,J)$-locally recoverable.

In Section \ref{QLRC} we study quantum stabilizer $(r,\delta)$-LRCs, $Q(C)$. That is, we consider stabilizer codes given by symplectic self-orthogonal $\mathbb{F}_q$-linear codes $C \subseteq \mathbb{F}_{q}^{2n}$; i.e., $C \subseteq C^{\perp_s}$, where duality is with respect to the symplectic form defined in (\ref{xpg6}). Considering the concepts of {\it shortening} of a classical code $D$, at the set $I$, denoted $\sigma_I(D)$, and {\it puncturing} of $D$ at $J$, denoted $\pi_J(D)$, Theorem \ref{IJ} gives a necessary and sufficient condition on the code $C$ and its symplectic dual for $Q(C)$ to be $(I,J)$-locally recoverable. Note that, contrary to usual, $I$ and $J$ refers to the coordinates of the codewords we keep (see the beginning of Section \ref{prel}). Theorem \ref{IJ} allows us to prove Propositions \ref{teo8} and \ref{teo11} which supply conditions implying or discarding $(I,J)$-local recoverability of $Q(C)$. These conditions depend only on the cardinality of $I$ and $J$ and certain generalized symplectic weights of the symplectic dual of $C$.

As a consequence of Theorem \ref{IJ}, we deduce the main result in the paper, {\it Theorem} \ref{quantumsyp}. It states that a quantum stabilizer code $Q(C) \subseteq \mathbb{C}^{q^n}$, defined by a symplectic self-orthogonal $q$-ary linear code $C \subseteq \mathbb{F}_q^{2n}$, is a quantum $(r,\delta)$-LRC if and only if
\[
\sigma_I\left[\pi_J \left(C^{\perp_s}\right)\right] = \sigma_I(C)
\]
for suitable subsets $J$ of cardinality $\leq r + \delta -1$ containing each index $i$ in $\{1, \ldots, n\}$ and for all subset $\emptyset \neq I \subsetneq J$ of cardinality $\delta -1$. 


Section \ref{La4} starts by specializing the results in Section \ref{QLRC} to QECCs coming from Hermitian or Euclidean self-orthogonal linear  codes $C$ included in $\mathbb{F}_{q^2}^{n}$ or $\mathbb{F}_{q}^{n}$, respectively. As can be expected, a set of  equalities as  in Theorem \ref{quantumsyp}, where one replaces symplectic  dual with Hermitian (respectively, Euclidean) dual, gives a necessary and sufficient condition for $Q(C)$ being an  $(r,\delta)$-LRC.

Replacing self-orthogonality of the involved classical codes with dual-containing, our second main result, {\it Theorem \ref{relation}}, proves that under the condition $\delta \leq d(C^{\perp_h})$ (respectively, $\delta \leq d(C^{\perp_e})$)
stabilizer $(r,\delta)$-LRCs, $Q'(C)$, and classical $(r,\delta)$-LRCs, $C$,  are equivalent objects. Here, $Q'(C)$ stands for the stabilizer code coming from a Hermitian  or Euclidean  dual-containing classical linear code $C$. The symbol $C^{\perp_h}$ (respectively, $C^{\perp_e}$) denotes the dual of $C$ with respect to the Hermitian $\cdot_h$ (respectively, Euclidean $\cdot_e$) inner product.


Then,
we propose in Theorem \ref{SingletonQ} the following Singleton-like bound for quantum stabilizer $(r,\delta)$-LRCs, $Q'(C)$, with parameters $[[n,k, \geq d(C)]]_q$:
\begin{equation}
\label{AA}
k + 2 d(C) + 2\left(\left\lceil\frac{n+k}{2r}\right\rceil-1\right)(\delta-1) \leq n+2.
\end{equation}

A quantum code $Q'(C)$, as above, is pure whenever its minimum distance coincides with $d(C)$ and, thus, a pure code $Q'(C)$ is named to be {\it optimal} if it attains the above bound. Making use of Theorems \ref{relation} and \ref{SingletonQ}, Section \ref{examples} contains some results implying optimality of certain families of pure quantum $(r,\delta)$-LRCs.

Section \ref{prel} contains some definitions and results which are needed for the understanding and development of this article.
At the end of this section we explain when a set of erasures is correctable by a quantum code.
Section \ref{QLRC} is the core of the paper, after introducing the concept of quantum $(r,\delta)$-LRC, we study  $(I,J)$-local recoverability of quantum stabilizer codes and, finally, in Subsection \ref{section33}, we state our main result. Section \ref{La4} treats quantum $(r,\delta)$-LRCs when they come from Hermitian or Euclidean inner  products. In Subsection \ref{subsect42} we state our second main result relating classical and quantum  $(r,\delta)$-LRCs. A Singleton-like bound is also given at the end of this section.
Subsection \ref{51} provides some families of optimal pure stabilizer $(r,\delta)$-LRCs coming from $\emptyset$-affine variety codes.  In Subsection \ref{52} we show that Hermitian dual-containing  $q^2$-ary MDS codes give rise to optimal pure stabilizer $(r,\delta)$-LRCs and that the same happens with the Euclidean case.  Finally, setting $\delta =2$, we find an optimal pure quantum $(r,2)$-LRC which, regarded as a quantum $r$-LRC, attains the bound (\ref{SingletonRQ}).

\section{Preliminaries}
\label{prel}

In this paper we use error-correcting linear codes $C \subseteq \mathbb{F}_q^n$ over finite fields $\mathbb{F}_q$ of cardinality $q=p^m$, $p$ being a prime number and $m$ a positive integer, and denote by $[n,k,d]_q$ their parameters. First we recall the concepts of shortening and puncturing of a code at a set $R$ because they will be useful; note that in our definitions, $R$ refers to the coordinates of the codewords we keep.

Set $R \subseteq \{1, \ldots, n\}$, assume that $r =  \mathrm{card}(R)$ and consider the projection map on the coordinates of $R$, $\pi_R: \mathbb{F}_q^n \rightarrow \mathbb{F}_q^r$. We define the {\it puncturing of $C$ at $R$} as the set $$\pi_R (C) = \{ \mathbf{c}_R :=(c_j)_{j \in R} \; :  \; \mathbf{c}=(c_1, \ldots, c_n) \in C\},$$ $\pi_R (C)$ is also named the {\it punctured code} of $C$ at $R$. We also define the {\it shortening of $C$ at $R$} as the set $$\sigma_R (C) = \{ \mathbf{c}_R\; : \; \mathbf{c}=(c_1, \ldots, c_n) \in C \mbox{ satisfying supp}(\mathbf{c}) \subseteq R \},$$
where supp$(\mathbf{c}) := \{j \in \{1, \ldots, n\} : c_j \neq 0\}$. We also say that $\sigma_R (C)$ is the {\it shortened code} of $C$ at $R$. By \cite{Pless}, it holds that
\begin{equation*}
  \label{dualpunct}
  \pi_R \left( C^{\perp_e} \right) = \left[ \sigma_R \left( C \right) \right]^{\perp_e}.
\end{equation*}

We divide the rest of this section into two subsections which recall concepts and facts which are necessary for our results. The first subsection deals with classical $(r,\delta)$-LRCs.

\subsection{Classical $(r,\delta)$-locally recoverable codes}
\label{LRC}



We start with the classical definition of LRC. Let $C \subseteq \mathbb{F}_q^n$ be a linear $q$-ary code as before and  set $i \in \{1, \ldots, n\}$ and $R \subseteq \{1, \ldots, n\}$ such that $i \notin R$. $R$ is said to be a {\it recovery set} for the $i$-th coordinate of $C$  whenever an erasure in the coordinate $c_i$ of a codeword $\mathbf{c}=(c_1, \ldots, c_n) \in C$ can be recovered from the coordinates of $\mathbf{c}$ with indices in $R$. This allows us to define the {\it locality of the $i$-th coordinate in $C$} as the smallest cardinality of a recovery set for the $i$-th coordinate and give the following definition:
\begin{de}
{\rm An error-correcting code $C$ is an $r$-LRC (or an LRC with locality $r$) whenever any coordinate of $C$ admits a recovery set, $r$ being the largest locality of its coordinates.}
\end{de}

The following twofold result is well-known. See \cite[Proposition 2.1]{GFMC} and \cite[Proposition 1.1]{GHM2020}.

\begin{pro}
\label{prop1}

With the above notation the following results hold.
\begin{enumerate}
\item A set of coordinates $R \subseteq \{1, \ldots, n\}$ is locally recoverable for the $i$-th coordinate  of a code $C$ of length $n$, $i \notin R$, if and only if the minimum (Hamming) distance $d$ of the code $\pi_{\bar{R}} (C)$, where $\bar{R}:= R \cup \{i\}$, satisfies $d[\pi_{\bar{R}} (C)] \geq 2$.
\item Let $r$ be the locality of an LRC, $C$, then $r \geq d(C^{\perp_e}) -1$, $C^{\perp_e}$ being the Euclidean dual of $C$.
\end{enumerate}

\end{pro}

As shown in \cite{GHSY2012}, parameters $[n,k,d]_q$ and locality $r$ of an LRC code satisfy a Singleton-like bound:

\begin{equation}
\label{Singleton1}
k+d+ \left\lceil{\frac{k}{r}} \right\rceil \leq n+2.
\end{equation}
Optimal LRCs are those meeting the bound in the above inequality. 

As shown in Proposition \ref{prop1}, LRCs correct only an erasure and thus, they are only useful when a node of the system fails. However simultaneous failures are not unusual. The so-called $(r,\delta)$-LRCs are designed to address this problem. 

\begin{de}
An error-correcting code $C \subseteq \mathbb{F}_q^n$ is an $(r,\delta)$-LRC (or an LRC with  locality $(r,\delta)$) if for each index $i \in \{1, \ldots, n\}$, there exists a set of indices $J \subseteq \{1, \ldots, n\}$ that contains $i$ and satisfies the following two conditions:
\begin{enumerate}
\item $\mathrm{card}\,(J) \leq r + \delta -1$.
\item $d[\pi_{J} (C)] \geq \delta$.
\end{enumerate}
\end{de}

The set $J$ is named an $(r,\delta)$-{\it recovery set} for the $i$-th coordinate  of $C$. Our definition implies that, for any set $I \subsetneq J$ of cardinality $\delta -1$, the coordinates corresponding to $I$ can be corrected from the knowledge of the coordinates corresponding to $J \setminus I$.

A Singleton-like bound has also been introduced for $(r,\delta)$-LRCs \cite{PKLK2012}. Let us state it.

\begin{pro}
\label{Singleton-2}
Let $C$ be an $(r,\delta)$-LRC with parameters $[n,k,d]_q$. Then, the following inequality holds.
\begin{equation}
\label{Singleform}
k+d+ \left(\left\lceil{\frac{k}{r}} \right\rceil -1 \right)\left( \delta -1 \right) \leq n+1.
\end{equation}
\end{pro}

As before, an optimal $(r,\delta)$-LRC is an $(r,\delta)$-LRC  that meets the bound in Proposition \ref{Singleton-2}. In this paper, we are interested in $(r,\delta)$-LRCs, therefore, unless otherwise specified, the phase optimal (classical) code will stand for an optimal $(r,\delta)$-LRC. 

We conclude this subsection with a result stated in \cite[Corollary 1.5]{GHM2020}
which extends that stated in Part (2) of Proposition \ref{prop1}. To do it we remind that, given an $[n,k,d]_q$ classical code $C$, for each index $1 \leq t \leq k$, the value
\[
\omega_t (C) := \min \{\mathrm{card}\,[\mathrm{supp}(C')] : C' \mbox{ is a $t$-dimensional subcode of } C \}
\]
is named the $t$-th {\it generalized Hamming weight of $C$} \cite{wei91,helleseth92}.

\begin{pro}
\label{prop3}
Let $C$ be an $(r,\delta)$-LRC. Then $r + \delta \geq \omega_{\delta-1} (C^{\perp_e}) +1$.
\end{pro}

Our last subsection briefly recalls basic information on quantum stabilizer codes.

\subsection{Quantum stabilizer codes}
\label{qquantum}

To introduce quantum stabilizer codes, we keep the above notation and use the definition and ideas given in \cite{AK}.
A linear character of a finite commutative group $S$ is a group homomorphism from $S$  to $\mathbb{C}^*$, where $\mathbb{C}^*$ denotes the multiplicative group of the complex numbers. Set $\xi = e^{\frac{\iota 2 \pi}{p}}$, $\iota = \sqrt{-1}$, a primitive $p$-th root of unity, and consider a nice error basis on $\mathbb{C}^{q^n}$, $\epsilon_n := \{ E_{(\mathbf{a},\mathbf{b})} \; : \; (\mathbf{a}|\mathbf{b}) \in \mathbb{F}_q^{2n}\}$, where $E_{(\mathbf{a},\mathbf{b})}$ are error operators obtained by tensoring generalized Pauli matrices acting on $\mathbb{C}^q$, as defined in \cite{AK}. Consider the group
\[
G_n := \{\xi^\ell E_{(\mathbf{a},\mathbf{b})} \; : \; (\mathbf{a}|\mathbf{b}) \in \mathbb{F}_q^{2n}, 0 \leq \ell \leq p-1\}.
\]
Then, to define a quantum stabilizer code, one needs to consider a commutative subgroup $S$ of $G_n$ with identity $\mathcal{I}$ and a linear character $\lambda$ of $S$ such that $\lambda (\xi \mathcal{I}) = \xi$; for convenience we assume that
$S$ contains the center $\{ \xi^\ell \mathcal{I} \; : \; 0 \leq \ell \leq p-1\}$
of $G_n$. Then, an (attached to $S$ and $\lambda$) {\it quantum stabilizer code} (or simply a stabilizer code) $Q$ is the common eigenspace
\[
Q = Q(S) := \bigcap_{E \in S} \{\mathbf{v} \in \mathbb{C}^{q^n} \; : \; E \mathbf{v} = \lambda(E) \mathbf{v} \}.
\]
Any code $Q$ as before is associated to a character but, unless necessary, we will not mention it.

Setting $\mathbf{a} = (a_1, \ldots, a_n)$ and $\mathbf{b} = (b_1, \ldots, b_n)$, $a_j, b_j \in \mathbb{F}_q$, the {\it symplectic weight} of an element $(\mathbf{a}|\mathbf{b}) \in \mathbb{F}_q^{2n}$ is defined as $$\mathrm{swt} (\mathbf{a}|\mathbf{b}) := \mathrm{card}\, \{ j : (a_j, b_j) \neq (0,0)\}.$$
In addition, the weight $\mathrm{wt}(E)$ of $ E = \xi^\ell E_{(\mathbf{a},\mathbf{b})}$ is defined as $\mathrm{swt} (\mathbf{a}|\mathbf{b})$. Finally, $Q$ has {\it minimum distance} $d$ if it detects all errors in $G_n$ of weight less than $d$ but not some of weight $d$. An $((n,K,d))_q$ quantum stabilizer code is a $K$-dimensional subspace of $\mathbb{C}^{q^n} =\mathbb{C}^q \otimes \cdots \otimes \mathbb{C}^q$ with minimum distance $d$. In case $K=q^k$, it is said to be an $[[n,k,d]]_q$ stabilizer code.

A quantum stabilizer code  defines a code of length $2n$
\[
C := \{ (\mathbf{a}| \varphi^{-1} \mathbf{b}) \; : \; \xi^\ell E_{(\mathbf{a},\mathbf{b})} \in S\},
\]
for a convenient automorphism $\varphi$ of $\mathbb{F}_q$ which is translated componentwise to $\mathbb{F}_q^n$. The automorphism $\varphi$ is chosen so that $C$ is self-orthogonal with respect to the following trace-symplectic form:
\[
(\mathbf{a}| \mathbf{b}) ._{\mathrm{trs}} (\mathbf{c}| \mathbf{d}) = \mathrm{tr}_{q/p} ( \mathbf{a} ._e \mathbf{d} - \mathbf{b} ._e \mathbf{c}),
\]
and we usually omit it.

Given a $q$-ary additive code $C \subseteq \mathbb{F}_q^n$, we denote by $C^{\perp_{\mathrm{trs}}}$ the dual code of $C$ with respect to the trace-symplectic form. The main theorem in this regard is the following one:

\begin{teo}
\label{simplec}
The existence of an $((n,K,d))_q$ stabilizer code is equivalent to that of an additive code $C \subseteq \mathbb{F}_q^{2n}$ with cardinality $q^n/K$, which is self-orthogonal with respect to the trace-symplectic form, and, whenever $K>1$, $d$ is the minimum symplectic weight of the elements in  $C^{\perp_{\mathrm{trs}}} \setminus C$. When $K=1$, there is no undetectable error.
\end{teo}

This result has two useful consequences that we state next. Firstly, recall that for $\mathbf{a}, \mathbf{b} \in \mathbb{F}_{q^2}^{n}$, the {\it Hermitian inner product} is defined as $\mathbf{a} \cdot_h \mathbf{b} := \sum_{j=1}^{n} a_j b_j^q$. Remind also that given a $q^2$-ary linear code $C$, $C^{\perp_h}$ denotes its dual with respect to $\cdot_h$. We stand $\mathrm{wt}$ for Hamming weight.
\begin{teo}
\label{resto}
The following two items hold:
\begin{enumerate}
\item The existence of an  $[n,k,d]_{q^2}$ linear code $C$ which is Hermitian dual-containing (that is $C^{\perp_h} \subseteq C$) implies that of an $[[n, 2k-n, \geq d]]_q$ quantum stabilizer  code.
\item The existence of two $q$-ary linear codes $C_j$ with parameters $[n,k_j,d_j]_q$, $1 \leq j \leq 2$, such that $C_2^{\perp_e} \subseteq C_1$ implies the existence of an $[[n, k_1+k_2-n,d]]_q$ quantum stabilizer code with $d= \min \{\mathrm{wt}(\mathbf{z}) : \mathbf{z} \in (C_1 \setminus C_2^{\perp_e}) \cup (C_2 \setminus C_1^{\perp_e})\}$.
\end{enumerate}
\end{teo}

When stating our results we have followed the terminology in \cite{Ketkar}, however our definition of stabilizer codes comes from \cite{AK}. There, the results are initially stated and proved over $\mathbb{F}_p$ and later extended to $\mathbb{F}_q$. Details of how to perform that extension can be found in \cite[Section 3.1]{QINP3} in the context of entanglement-assisted codes. In this paper we will use this procedure.\\

We conclude this subsection by explaining when a set of erasures is correctable by a quantum code $Q$. Stand $A^\dagger$ for the conjugate transpose (or Hermitian conjugate) of a complex matrix $A$. As before, denote by $\epsilon_1$
the set of (generalized) Pauli matrices acting on $\mathbb{C}^q$ and, for a  $q \times q$ complex matrix $\rho$, consider the completely depolarizing channel on a single qudit in $\mathbb{C}^q$:
\[
\Gamma(\rho) = \frac{1}{q^2} \sum_{E \in \epsilon_1} E \rho E^\dagger.
\]
Note that, with $\Gamma$, every possible error in $\epsilon_1$ (including the $q \times q$ identity matrix $\mathcal{I}_q$)
is applied with probability $1/q^2$. Consider now a quantum code $Q \subseteq \mathbb{C}^{q^n}= \mathbb{C}^q \otimes \cdots \otimes \mathbb{C}^q$ and a set of indices $I \subseteq \{1, \ldots, n\}$. Set $\Gamma^I$ the map $\bigotimes_{j=1}^n \Gamma_j$ where $\Gamma_j = \Gamma$ when $j \in I$ and $\Gamma_j = \mathcal{I}_{q}$ otherwise; that is the map applying $\Gamma$ on all qudits at (the positions in) $I$. Then, by definition, {\it erasures at $I$  are correctable by $Q$} if there exists a recovery operator $\mathcal{R}_{Q,I}$, which is a trace-preserving quantum operation (in the sense of \cite[Chapter 8]{chuangnielsen}),
such that
\begin{equation}
  \mathcal{R}_{Q,I} \circ \Gamma^I (\ket{\varphi}\bra{\varphi}) = \ket{\varphi}\bra{\varphi} \label{eq1}
\end{equation}
for all $\ket{\varphi} \in Q$.

\begin{rem}
An alternative definition, perhaps more popular than the previous one, states that a quantum code corrects erasures at $I$ whenever its decoder can correct any error $E \in \epsilon_n$ by assuming that the support of $E$ is contained in $I$.

The above alternative definition and ours are equivalent. Indeed, if a code can perfectly (meaning quantum fidelity one) correct a Pauli error supported by $I$ happening with equal probability, the code can correct every unknown Pauli error supported by $I$.
Conversely, if a code can perfectly correct every unknown Pauli error supported by $I$, the code can also correct those errors happening with equal probability. Moreover, while erasure correction assumes decoder's knowledge of $I$, knowledge of $I$ does not change the equivalence in this paragraph.
\end{rem}



\section{Quantum $(r,\delta)$-locally recoverable codes}
\label{QLRC}

\subsection{Definitions}

We start this section by introducing the concept of quantum code $Q \subseteq \mathbb{C}^{q^n}$ with alphabet size $q$ that corrects  erasures at the positions in $I \subseteq \{1, \ldots, n\}$  from the qudits at a set $J \supsetneq I$, $J \subseteq \{1, \ldots, n\}$. 
Recall the definition of $\Gamma^I$ given above.

\begin{de}
\label{I-J}
Given two sets of indices as before $\emptyset \neq I \subsetneq J \subseteq \{1, \ldots, n\}$, a quantum code  $Q \subseteq \mathbb{C}^{q^n}$ is said to be an {\it $(I,J)$-locally recoverable code} if there exists a trace-preserving quantum operation $\mathcal{R}_{Q,I}^{J}$, which acts only on the qudits corresponding to $J$ and keeps untouched the remaining ones, such that
\begin{equation}
  \mathcal{R}_{Q,I}^J \circ \Gamma^I (\ket{\varphi}\bra{\varphi}) = \ket{\varphi}\bra{\varphi} \label{eq2}
\end{equation}
for all $\ket{\varphi} \in Q$.

\end{de}

This definition allows us to introduce the concept of quantum $(r,\delta)$-locally recoverable code.

\begin{de}
\label{lcr}
A quantum code  $Q \subseteq \mathbb{C}^{q^n}$ is said to be a {\it quantum $(r,\delta)$-locally recoverable code} (quantum $(r,\delta)$-LRC) if for each index $i \in \{1, \ldots, n\}$, there exists a set $J \subseteq \{1, \ldots, n\}$ containing $i$ with $\mathrm{card}\,(J) \leq r + \delta-1$ satisfying that for all subset $I \subseteq J$ of cardinality $\delta -1$, $Q$ is $(I,J)$-locally recoverable.
\end{de}

\subsection{$(I,J)$-local recoverability for quantum stabilizer codes}
\label{section32}
In this paper, for constructing stabilizer codes we consider $\mathbb{F}_q$-linear codes and therefore a slightly more restrictive version of the trace-symplectic form. Indeed, for $(\mathbf{a}|\mathbf{b}), (\mathbf{c}|\mathbf{d}) \in \mathbb{F}_q^{2n}$, we consider the following {\it symplectic form}:
\begin{equation}
\label{xpg6}
(\mathbf{a}|\mathbf{b}) \cdot_s (\mathbf{c}|\mathbf{d}) := \mathbf{a} \cdot_e \mathbf{d} - \mathbf{b} \cdot_e \mathbf{c}.
\end{equation}

From now on, $Q(C)$ denotes the quantum stabilizer code defined by a symplectic self-orthogonal (with respect to $\cdot_s$) classical  $\mathbb{F}_q$-linear code $C \subseteq \mathbb{F}_q^{2n}$; that is $C \subseteq C ^{\perp_s}$, where the duality is with respect the symplectic form $\cdot_s$.

We start by explaining a standard procedure for correcting erasures at $I \subsetneq \{1, \ldots, n\}$ of the code $Q(C)$.
Our procedure is the same for codes $C \subseteq \mathbb{F}_{q^2}^{n}$ (respectively, $C \subseteq \mathbb{F}_{q}^{n}$) such that $C \subseteq C^{\perp_h}$ (respectively, $C \subseteq C^{\perp_e}$).

Let $[[n,k]]_q$ be the parameters of $Q(C)$, then $\dim C= n-k$ and $\dim (C^{\perp_s}) = n + k$. Consider a basis $\{\mathbf{h}_1, \ldots, \mathbf{h}_{n-k} \}$ of the linear space $C$ and the surjective map
\[
f: \mathbb{F}_q^{2n} \rightarrow \mathbb{F}_q^{n-k}; \; \;  f(\mathbf{y}) = (\mathbf{y} \cdot_s \mathbf{h}_1, \ldots, \mathbf{y} \cdot_s \mathbf{h}_{n-k})
\]
whose kernel is $C^{\perp_s}$. Then, in a {\it first step,} for each index $1 \leq \ell \leq n-k$, the decoder measures an observable whose eigenspaces are the same as $E_{\mathbf{h}_\ell}$ and the measurement outcomes provide the vector $(s_1, \ldots, s_{n-k}) \in \mathbb{F}_q^{n-k}$, where $s_\ell = \mathbf{e} \cdot_s \mathbf{h}_\ell$, for some error vector $\mathbf{e} \in \mathbb{F}_q^{2n}$. In a {\it second step,} setting $\mathbf{y} = (\mathbf{a}, \mathbf{b})$ an unknown, the decoder considers the  linear system of equations \begin{equation}
\label{system}
f(\mathbf{y})= (s_1, \ldots, s_{n-k}); \; a_j = b_j = 0, j \not \in I.
\end{equation}
{\it Finally,} since we desire to correct erasures at $I$,  if the solution $\mathbf{e}$ of the system is unique modulo $C$, then the decoder applies the inverse $E_{\mathbf{e}}^{-1}$. Otherwise, the decoder declares failure.

We take into account the above procedure, to state the following result.

\begin{teo}
\label{twostars}
Let $C$ be a  symplectic self-orthogonal classical $\mathbb{F}_q$-linear code and $Q(C)$ the corresponding stabilizer code. Keep also the above notation. Then, the equality
\begin{equation}
\label{correrasur}
\sigma_I(C) = \sigma_I (C^{\perp_s})
\end{equation}
is a necessary and sufficient condition for $Q(C)$ to correct erasures at $I$.
\end{teo}

\begin{proof}
By \cite[Equation (3)]{matsumoto17uni}, the system of linear equations (\ref{system}) has a unique solution when Equality (\ref{correrasur}) holds, and therefore (\ref{correrasur}) implies that $Q(C)$ corrects erasures at $I$.

To conclude the proof, we are going to show that if Equality (\ref{correrasur}) is false, then no decoding procedure can correct erasures at $I$. Indeed, if Equality (\ref{correrasur}) does not hold, there is an erasure $\mathbf{e} = (\mathbf{a}, \mathbf{b}) \in C^{\perp_s} \setminus C$ with $a_j=b_j=0$ for $j \not \in I$. Then $E_{\mathbf{e}}$ sends some quantum codeword in $Q(C)$ to another one in $Q(C)$, which prevents to correct all erasures at $I$ by any decoding procedure (in the sense of (\ref{eq1})).
\end{proof}

In the rest of this subsection we study the case $J \subsetneq \{1, \ldots, n\}$. We divide our study in two parts. The first one provides a sufficient condition for a stabilizer code to be $(I,J)$-locally recoverable. In the second part, we will show that the condition stated in the first part is also necessary.

As before mentioned, we state and prove our results for the case when the alphabet size is a prime number $p$; this is because they can be extended to a general $q$ following the reasoning given in \cite[Section 2.1]{QINP3}.

\subsubsection{A sufficient condition}
\label{sufficient}

We desire to correct erasures from qudits at a set $J \subsetneq \{1, \ldots, n\}$, therefore we are interested in what happens when considering codewords in $Q(C)$ and discarding their  qudits at $\overline{J} = \{1, \ldots, n\} \setminus J$. Thus let $\ket{\varphi} \in Q(C)$ be a codeword in our stabilizer code, the partial trace over $\overline{J}$, $\sigma = \mathrm{tr}_{\overline{J}}[\ket{\varphi}\bra{\varphi}]$, of the density matrix $\ket{\varphi}\bra{\varphi}$ allows us to consider the part of the codeword $\ket{\varphi}$ in $J$.

In our case, where  $C \subseteq \mathbb{F}_p^{2n}$, for a vector $ \mathbf{y}= (\mathbf{a}|\mathbf{b})$ we denote $\pi_J(\mathbf{y}) = (a_j | b_j)_{j \in J}$ and, as before, the {\it shortening of $C$ at} $J$ is defined as $$\sigma_J (C) = \{ \pi_J(\mathbf{y}) \; : \; \mathbf{y} \in C \mbox{ and } \mathrm{supp}(\mathbf{a}), \mathrm{supp}(\mathbf{b})\subseteq J \}.$$ Analogously, the {\it puncturing of $C$ at} $J$ is $\pi_J(C) =  \{ \pi_J(\mathbf{y}) \; : \; \mathbf{y} \in C \}$. Then, it is clear that $\ket{\varphi}$ is an eigenvector of an observable $E_{\mathbf{y}}$, where $\mathbf{y}$ is given by a vector $\mathbf{z} \in \sigma_J(C)$ to which we add zeroes at the coordinates supported in $\overline{J}$. Our first result relates the column space of the above density matrix, $\mathrm{col}(\sigma)$, to the stabilizer code defined by the shortened code $\sigma_J (C)$.

\begin{pro}
\label{era3-13}
Let $Q(C) \subseteq \mathbb{C}^{p^{n}}$ be the quantum stabilizer code defined by a symplectic self-orthogonal classical code $C$. Let $J \subsetneq \{1, \ldots, n\}$ be a set of indices. Consider an element $\ket{\varphi} \in Q(C)$ and the density matrix $\sigma = \mathrm{tr}_{\overline{J}}[\ket{\varphi}\bra{\varphi}]$. Set $Q(\sigma_J (C)) \subseteq \mathbb{C}^{p^{\mathrm{card}(J)}}$ the stabilizer code defined by the shortened code $\sigma_J (C)$ such that, for each pair of codewords $\mathbf{y}\in C$ and $\mathbf{z} \in \sigma_J (C)$ behaving as indicated before the statement, with respective observables $E_\mathbf{y}$ and $E_\mathbf{z}$, the characters $\lambda(E_\mathbf{y})$ and $\lambda(E_\mathbf{z})$ coincide. Then, the column space, $\mathrm{col}(\sigma)$, satisfies
\[
\mathrm{col}(\sigma) \subseteq Q(\sigma_J (C)).
\]
\end{pro}

\begin{proof}
It is clear that $\sigma_J (C) \subseteq \mathbb{F}_p^{2 \, \mathrm{card}(J)}$.
Let $V \subset \mathbb{C}^{p^n}$ be the eigenspace associated to the eigenvalue $\lambda(E_{\mathbf{y}})$ of $E_{\mathbf{y}}$. Denote by $P(\mathbf{y})$  the orthogonal projection from $\mathbb{C}^{p^n}$ onto $V$. Consider an observable $A$ on $\mathbb{C}^{p^n}$ whose eigenspaces are the same as those of $E_{\mathbf{y}}$. Outcomes by measuring $A$ can be identified with eigenvalues of $E_{\mathbf{y}}$. Set $\mathrm{tr}$ the trace map, it is an axiom of quantum theory  that the probability of getting a measurement outcome $\lambda(E_{\mathbf{y}})$ is $\mathrm{tr}[\ket{\varphi}\bra{\varphi} P(\mathbf{y})]$ when the measured system is in pure state $\ket{\varphi}$ \cite{chuangnielsen}.
Then, the probability of not getting $\lambda(E_{\mathbf{y}})$ equals
\[
1-\mathrm{tr}[\ket{\varphi}\bra{\varphi} P(\mathbf{y})] =
1-\mathrm{tr}[\sigma P(\mathbf{z})] = \mathrm{tr}[\sigma (\mathcal{I}_{p^{\mathrm{card}(J)}}-P(\mathbf{z}))]=0,
\]
where
$\mathcal{I}_{p^{\mathrm{card}(J)}}$ is the identity matrix of size $p^{\mathrm{card(J)}} \times p^{\mathrm{card(J)}}$ and $P(\mathbf{z})$ is the orthogonal projection from $\mathbb{C}^{p^{\mathrm{card}(J)}}$ onto the eigenspace associated to the eigenvalue $\lambda(E_{\mathbf{z}})$ of $E_{\mathbf{z}}$. As a consequence, and since $\sigma$ is positive semidefinite, one gets that $\mathrm{col}(\sigma) \subseteq \mathrm{col}(P(\mathbf{z}))$.

The relation between $\mathbf{z}$ and  $\mathbf{y}$ and the fact that $Q(\sigma_J (C))$ is the intersection of column spaces  $\mathrm{col}(P(\mathbf{z}))$, where $\mathbf{z}$ runs over $\sigma_J (C)$, concludes the proof.
\end{proof}


Let us state our main result in this subsection.

\begin{pro}
\label{th12}
Let $Q(C) \subseteq \mathbb{C}^{p^n}$ be a stabilizer code as above and consider sets of indices  $\emptyset \neq I \subsetneq J \subsetneq \{1, \ldots, n\}$. The codes' equality
\begin{equation}
\label{NAS}
\sigma_I\left[\pi_J \left(C^{\perp_s}\right)\right] = \sigma_I(C)
\end{equation}
is a sufficient condition for $Q(C)$ to be $(I,J)$-locally recoverable.
\end{pro}

\begin{proof}
Consider a codeword $\ket{\varphi} \in Q(C)$ for which qudits at $I$  were erased. We desire to recover  $\ket{\varphi}$ (in fact, the qudits at $I$) by measurements and a unitary operator on qudits at $J$. By Proposition \ref{era3-13}, every vector in $\mathrm{col}(\sigma)$, $\sigma = \mathrm{tr}_{\overline{J}}[\ket{\varphi}\bra{\varphi}]$, is a codeword in $Q(\sigma_J(C))$. Then, to recover erasures at $I$, we can use the above defined decoder $\mathcal{R}_{Q,I}$ (for the quantum code $Q = Q(\sigma_J(C))$). By Theorem \ref{twostars}, $Q(\sigma_J(C))$ corrects erasures at $I$ if and only if
\[
\sigma_I \left[ \left( \sigma_J(C) \right)^{\perp_s} \right] = \sigma_I \left[ \sigma_J(C) \right],
\]
which is equivalent to
\[
\sigma_I\left[\left(\sigma_J(C)\right)^{\perp_s}\right] =  \sigma_I(C),
\]
and, by \cite{Pless}, this last equality is equivalent to
\[
\sigma_I\left[\pi_J \left(C^{\perp_s}\right)\right] = \sigma_I(C).
\]
This concludes the proof.
\end{proof}

\subsubsection{A necessary and sufficient condition for $(I,J)$-local recoverability}
The aim of this section is to show that Equality (\ref{NAS}) is not only a sufficient but also a necessary condition for $Q(C)$ to be an $(I,J)$-locally recoverable code. Our proof will be a consequence of the forthcoming Lemmas \ref{elA}, \ref{lemma4} and \ref{elC}.

Our first result shows the necessity of Equality (\ref{NAS}) as long as certain requirement holds. Before to state it, we recall that a {\it completely mixed state} on an $\ell$-dimensional linear space $V \subseteq \mathbb{C}^{p^n}$ is an equal probabilistic mixture of the quantum states of an orthonormal basis of $V$.

\begin{lem}
\label{elA}
Let $Q(C) \subseteq \mathbb{C}^{p^n}$ be a stabilizer code given by a symplectic self-orthogonal $p$-ary code $C \subseteq \mathbb{F}_p^{2n}$, and consider sets of indices $\emptyset \neq I \subsetneq J \subsetneq \{1, \ldots, n\}$. Set $\overline{J} = \{1, \ldots, n\} \setminus J$. Suppose that there exists an environment $\mathcal{E}$ surrounding $\mathbb{C}^{p^n}$ and an entangled quantum codeword $\ket{\Phi} \in Q(C) \otimes \mathcal{E}$
such that the partial trace of $\ket{\Phi}\bra{\Phi}$ over $\overline{J}$ and $\mathcal{E}$ is a completely mixed state on $Q(\sigma_J(C))$. Then the codes' equality stated in (\ref{NAS}):
\[
\sigma_I\left[\pi_J \left(C^{\perp_s}\right)\right] = \sigma_I(C)
\]
is a necessary condition for $Q(C)$ to be $(I,J)$-locally recoverable.
\end{lem}
\begin{proof}
We reason by contradiction and thus we suppose that Equality (\ref{NAS}) is false.
Then, there exists some element $\mathbf{e} \in \pi_J(C^{\perp_s})$ with zero coordinates out of
$I$ such that $\pi_I(\mathbf{e}) \notin \sigma_I(C)$. Then $E_{\mathbf{e}} \ket{\xi} \in Q(\sigma_J(C))$
for all $\ket{\xi} \in Q(\sigma_J(C))$. Since $\mathbf{e}$ is not in the stabilizer of $Q(\sigma_J(C))$,
there exists  $\ket{\zeta} \in Q(\sigma_J(C))$ such that $E_{\mathbf{e}} \ket{\zeta}$ is not a scalar multiple of $\ket{\zeta}$ and it is physically different from $\ket{\zeta}$.

Suppose that a local recovery procedure $\mathcal{S} := \mathcal{S}^J_{Q(\sigma_J(C)), I}$ keeps $E_{\mathbf{e}} \ket{\zeta}$ when no erasure happens. Then $\mathcal{S}$ does not correct $\ket{\zeta}$  when the erasure $\mathbf{e}$ happens
as $\mathcal{S}$ keeps $E_{\mathbf{e}} \ket{\zeta}$ the same.

Consider the completely mixed quantum state on $Q(\sigma_J(C))$
$$\rho = \frac{1}{\dim  Q(\sigma_J(C))} [\ket{\zeta}\bra{\zeta}  + \sum_{\mu \in U} \ket{\mu}\bra{\mu} ],$$
where $\{\ket{\zeta} \} \cup \{ \ket{\mu} : \mu \in U \}$ forms an orthonormal basis of $Q(\sigma_J(C))$. Since $\ket{\zeta}$  is uncorrectable, the ensemble average fidelity \cite[(9.127)]{chuangnielsen} between
$\rho$ and $[\mathcal{S}(\rho \mbox{ after  erasure } \mathbf{e})]$ is not one (see also \cite[Exercise 9.23]{chuangnielsen}). This fact, that of $\rho$ is the partial trace of $\ket{\Phi}\bra{\Phi}$ over $\overline{J}$ and $\mathcal{E}$ and Equalities (58) in \cite{schumacher96} show that the fidelity between $\ket{\Phi}\bra{\Phi}$ and
$\left[\mathcal{S} \otimes \mathcal{I}_{\overline{J}, \mathcal{E}} \left( \ket{\Phi}\bra{\Phi} \mbox{ after erasure } \mathbf{e} \right)\right]$ is strictly less than one.

By \cite[Theorem V.3]{knill97}, the minimum pure state fidelity for $Q(C)$ and $\mathcal{S}$ is not one.
Therefore, for every recovery procedure $\mathcal{R}^J_{Q(C), I}$, the minimum pure state fidelity neither is one.
This means that there is no local recovery procedure for $J$ with erasures at $I$, which is a contradiction and it finishes the proof.
\end{proof}

Our next result proves that the requirement indicated in Lemma \ref{elA} holds whenever $C$ is symplectic self-dual and $\mathcal{E}= \mathbb{C}$. We note that given a code $C \subseteq C^{\perp_s} \subseteq \mathbb{F}_p^{2n}$, by \cite[Section 20]{aschbacher00}, one can always find a $p$-ary code $C_{\mathrm{max}}$ such that
\begin{equation}
\label{max}
C \subseteq C_{\mathrm{max}} = C_{\mathrm{max}}^{\perp_s} \subseteq C^{\perp_s}.
\end{equation} 

\begin{lem}
\label{lemma4}
Let $C$ be a  symplectic self-dual $p$-ary classical code, that is $C = C^{\perp_s} \subseteq \mathbb{F}_p^{2n}$. Let $J \subsetneq \{1, \ldots, n\}$ be a nonempty set of indices and set $\overline{J} = \{1, \ldots, n\} \setminus J$.  Then, for each $\ket{\varphi} \in Q(C)$, $\mathrm{tr}_{\overline{J}}[\ket{\varphi}\bra{\varphi}]$ is a completely mixed state on $Q(\sigma_J(C))$.
\end{lem}

\begin{proof}
Let $\ket{\varphi}$ be as in the statement; at the end of the proof we will explicitly compute the partial trace over $\overline{J}$
of $\ket{\varphi}\bra{\varphi}$, which will allow us  to prove the result. We start by constructing two symplectic self-dual codes included in $\mathbb{F}_p^{2 \, \mathrm{card}\,(J)}$, $\mathbb{F}_p^{2 \, (n-\mathrm{card}\,(J))}$, respectively, which we denote by $G$ and $H$. 

Consider the code $\sigma_J(C)$ and a code $(\sigma_J(C))_{\mathrm{max}}$  as described in (\ref{max}). Then, $\sigma_J(C) \subseteq (\sigma_J(C))_{\mathrm{max}} = ((\sigma_J(C))_{\mathrm{max}})^{\perp_s} \subseteq \pi_J(C^{\perp_s}) =\pi_J(C)$. For simplicity,  denote $G:= (\sigma_J(C))_{\mathrm{max}}$. Define
$$
D_1 : = \{\mathbf{x}\in C \;:\; \pi_J(\mathbf{x}) \in G \}
$$
and, then,
\begin{equation}
\label{jbar}
\pi_{\overline{J}} (D_1) \subseteq (\pi_{\overline{J}} (D_1))^{\perp_s}.
\end{equation}
Again, to simplify our notation, set $H:= \pi_{\overline{J}} (D_1)$. To prove (\ref{jbar}), let $\pi_{\overline{J}}(\mathbf{x}), \pi_{\overline{J}}(\mathbf{y}) \in H$, then the facts that $\mathbf{x}, \mathbf{y} \in C = C^{\perp_s}$ and $\pi_{J}(\mathbf{x}), \pi_{J}(\mathbf{y}) \in G = G^{\perp_s}$ prove the symplectic orhogonality of $\pi_{\overline{J}}(\mathbf{x})$ and $\pi_{\overline{J}}(\mathbf{y})$, and therefore Inclusion (\ref{jbar}).

The following sequence of equalities relates $\dim D_1$ to the cardinality of $\overline{J}$ and the dimension of $\sigma_J(C)$.
\begin{eqnarray*}
\label{eq102}
    && \dim D_1 \\
 &=& \dim C - [(\underbrace{\dim C - \dim \sigma_{\overline{J}}(C)}_{=\;\dim \pi_J(C)}) - (\underbrace{\dim D_1 - \dim \sigma_{\overline{J}}(C)}_{= \; \dim G}]\\
    &=& \dim C - (\underbrace{
    \dim \pi_J(C) - \dim G}_{ = \; \mathrm{card}\, (J) -\dim (\sigma_J(C)) }) \\
    &=& \mathrm{card}\,(\overline{J}) + \dim \sigma_J(C).
  \end{eqnarray*}
Let us explain the above equalities. Set $\mathbf{0}_{J}$ the zero vector with coordinates corresponding to $J$.  In the first equality, the first (respectively, second) equality under the curly brackets comes from the exact sequence $0 \rightarrow \mathbf{0}_{J} \bigoplus \sigma_{\overline{J}} (C) \rightarrow C \xrightarrow{\pi_J} \pi_J(C) \rightarrow 0$ (respectively, $0 \rightarrow \mathbf{0}_{J} \bigoplus \sigma_{\overline{J}} (C) \rightarrow D_1 \xrightarrow{\pi_J} G \rightarrow 0$); the fact that the kernel in the second exact sequence coincides with $\mathbf{0}_{J} \bigoplus \sigma_{\overline{J}} (C)$ holds by the definition of $D_1$.

The equality under the curly brackets in the second equality follows, on the one hand, from the chain of equalities $[\pi_J(C)]^{\perp_s}= \sigma_J (C^{\perp_s}) = \sigma_J(C)$ (see \cite{QINP3}) and from the fact that  $C$ is self-dual, which proves that $\dim \pi_J(C) = 2 \, \mathrm{card}\,(J) - \dim (\sigma_J(C))$; and, on the other hand, from the equality $\dim (G) = \mathrm{card}\,(J)$, which is true because $G$ is self-dual.

Finally, the last equality is clear by noting that $C$ is self-dual and therefore $\dim C = \mathrm{card}\,(J) + \mathrm{card}\,(\overline{J})$.

Now,  
by the definition of $D_1$ and $H$,
one has the exact sequence $0 \rightarrow \mathbf{0}_{\overline{J}} \bigoplus \sigma_{J} (C) \rightarrow D_1 \xrightarrow{\pi_{\overline{J}}} H \rightarrow 0$ and therefore, $\dim H = \dim D_1  - \dim \sigma_{J} (C) = \mathrm{card}\,(\overline{J})$ which, together with (\ref{jbar}), proves that $H:= \pi_{\overline{J}} (D_1) \subseteq \mathbb{F}_p^{
2 \, \mathrm{card}\,(\overline{J})} $ is self-dual. Recall that $G= (\sigma_J(C))_{\mathrm{max}}$.


Next we give an orthonormal basis of $Q(\sigma_J(C))$. Take $\ket{\varphi_J} \in Q(G) \subset Q(\sigma_J(C))$, then,  for all $\mathbf{v} \in \sigma_J(C) \subseteq G$, $E_{\mathbf{v}}\ket{\varphi_J} = \lambda(E_{\mathbf{v}})\ket{\varphi_J}$. Let $D_2$ be an $\mathbb{F}_p$-linear space  such that $ C = D_1 \bigoplus D_2$, then $\dim D_2 = \mathrm{card}\,(J) -\dim \sigma_J(C)$ by the previous sequence of equalities.

From our definitions, one has that $D_1 = C \cap \pi_J^{-1}(G)$ and $D_1 = C \cap \pi_{\overline{J}}^{-1}(H)$. Then, picking $\mathbf{x}\neq \mathbf{y} \in D_2$, one gets that $\pi_J(\mathbf{x}) + G \neq \pi_J(\mathbf{y}) + G$.
Therefore, $E_{\pi_J(\mathbf{x})}\ket{\varphi_J}$
is orthogonal to $E_{\pi_J(\mathbf{y})}\ket{\varphi_J}$.
Finally, the fact that $\dim D_2 = \mathrm{card}\,(J) -\dim \sigma_J(C) = \log_p \dim Q(\sigma_J(C))$ allows us to conclude that
\begin{equation}
\{E_{\pi_J(\mathbf{x})}\ket{\varphi_J} \; : \; \mathbf{x} \in D_2 \} \label{eq1001}
\end{equation}
is an orthonormal basis of the complex linear space $Q(\sigma_J(C))$.


Let $\ket{\varphi} \in Q(C)$, our next step is to give an expression for $\ket{\varphi}$ which will help in our proof. Consider the set of projections
$\left\{ E_{\pi_J(\mathbf{x})} \ket{\varphi_J}\bra{\varphi_J}E^\dagger_{\pi_J(\mathbf{x})}\otimes \mathcal{I}_{\overline{J}}: \mathbf{x} \in D_2 \right\}$, where $\mathcal{I}_{\overline{J}}$ is the identity matrix on the quantum symbols indexed by $\overline{J}$. The sum of these projections is equal to the identity mapping
on $Q(\sigma_J(C)) \otimes (\mathbb{C}^p)^{\otimes \left(n-\mathrm{card}(J)\right)}$.
Therefore, there exists $\mathbf{x}_0 \in D_2$ such that
$(E_{\pi_J(\mathbf{x}_0)} \ket{\varphi_J}\bra{\varphi_J}E^\dagger_{\pi_J(\mathbf{x}_0)}\otimes \mathcal{I}_{\overline{J}}) \ket{\varphi}$ is nonzero.
Set $\ket{\psi_J}:=E_{\pi_J(\mathbf{x}_0)} \ket{\varphi_J}$,
and
define $\psi_{\overline{J}}(\mathbf{0})$ by
\[
(\ket{\psi_J}\bra{\psi_J}\otimes \mathcal{I}_{\overline{J}}) \ket{\varphi}
= \frac{1}{\sqrt{\mathrm{card}\,(D_2)}}\ket{\psi_J} \otimes \ket{\psi_{\overline{J}}(\mathbf{0})},
\]
which implies that $\ket{\psi_{\overline{J}}(\mathbf{0})}$ is nonzero.

For $\mathbf{x} \in D_2$, define $\ket{\psi_{\overline{J}}(\mathbf{x})}$ by
\[
(E_{\pi_J(\mathbf{x})} \ket{\psi_J}\bra{\psi_J}E^\dagger_{\pi_J(\mathbf{x})}\otimes \mathcal{I}_{\overline{J}}) \ket{\varphi}
= \frac{1}{\sqrt{\mathrm{card}\,(D_2)}}E_{\pi_J(\mathbf{x})}\ket{\psi_J} \otimes \ket{\psi_{\overline{J}}(\mathbf{x})}.
\]
Recall that
the sum of the projections defined at the beginning of the previous paragraph is equal to the identity mapping, and thus we have the announced expression for $\ket{\varphi}$:
\begin{equation}
\ket{\varphi} = \sum_{\mathbf{x} \in D_2}\frac{1}{\sqrt{\mathrm{card}\,(D_2)}}E_{\pi_J(\mathbf{x})}\ket{\psi_J} \otimes \ket{\psi_{\overline{J}}(\mathbf{x})}. \label{eq101}
\end{equation}

Let $\mathbf{y}\in D_1$, then
\begin{eqnarray}
\label{newlemma15}
 \notag \lambda(E_{\mathbf{y}})\ket{\varphi} &=& \lambda(E_{\mathbf{y}}) \sum_{\mathbf{x} \in D_2}\frac{1}{\sqrt{\mathrm{card}\,(D_2)}}E_{\pi_J(\mathbf{x})}\ket{\psi_J} \otimes \ket{\psi_{\overline{J}}(\mathbf{x})}\\ \notag
  &=& E_{\mathbf{y}}\sum_{\mathbf{x} \in D_2}\frac{1}{\sqrt{\mathrm{card}\,(D_2)}}E_{\pi_J(\mathbf{x})}\ket{\psi_J} \otimes \ket{\psi_{\overline{J}}(\mathbf{x})}\\
  &=&
  \sum_{\mathbf{x} \in D_2}\frac{1}{\sqrt{\mathrm{card}\,(D_2)}}E_{\pi_J(\mathbf{y})}E_{\pi_J(\mathbf{x})}\ket{\psi_J} \otimes E_{\pi_{\overline{J}}(\mathbf{y})}\ket{\psi_{\overline{J}}(\mathbf{x})}.\\ \notag
&=&  \sum_{\mathbf{x} \in D_2}\frac{1}{\sqrt{\mathrm{card}\,(D_2)}}E_{\pi_J(\mathbf{x})}E_{\pi_J(\mathbf{y})}
\xi^{\pi_J(\mathbf{x}) \cdot_s  \pi_J(\mathbf{y})}\ket{\psi_J} \otimes E_{\pi_{\overline{J}}(\mathbf{y})}\ket{\psi_{\overline{J}}(\mathbf{x})}\\ \notag
&=&  \sum_{\mathbf{x} \in D_2}\frac{1}{\sqrt{\mathrm{card}\,(D_2)}}E_{\pi_J(\mathbf{x})}\eta(E_{\pi_J(\mathbf{y})})\xi^{\pi_J(\mathbf{x}) \cdot_s  \pi_J(\mathbf{y})}\ket{\psi_J} \otimes E_{\pi_{\overline{J}}(\mathbf{y})}\ket{\psi_{\overline{J}}(\mathbf{x})},
\end{eqnarray}
where $\xi$ is as defined in Subsection \ref{qquantum} and $\eta(E_{\pi_J(\mathbf{y})})$ is the eigenvalue of $E_{\pi_J(\mathbf{y})}$ to which
$\ket{\psi_J}$ belongs to.

Then we have
$E_{\pi_J(\mathbf{y})}E_{\pi_J(\mathbf{x})}\ket{\psi_J} = \eta(E_{\pi_J(\mathbf{y})})\xi^{\pi_J(\mathbf{x}) \cdot_s  \pi_J(\mathbf{y})}E_{\pi_J(\mathbf{x})}\ket{\psi_J}$ and, therefore, for $\mathbf{x} \in D_2$ and $\mathbf{y}\in D_1$, by comparing the right hand of the first and the last equalities in (\ref{newlemma15}), one has the equality
\[
\lambda(E_{\mathbf{y}}) E_{\pi_J(\mathbf{x})}\ket{\psi_J} \otimes \ket{\psi_{\overline{J}}(\mathbf{x})}
= \eta(E_{\pi_J(\mathbf{y})})\xi^{\pi_J(\mathbf{x}) \cdot_s  \pi_J(\mathbf{y})}E_{\pi_J(\mathbf{x})}\ket{\psi_J} \otimes
E_{\pi_{\overline{J}}(\mathbf{y})}\ket{\psi_{\overline{J}}(\mathbf{x})},
\]
which implies
\[
E_{\pi_{\overline{J}}(\mathbf{y})}\ket{\psi_{\overline{J}}(\mathbf{x})}
= \frac{\lambda(E_{\mathbf{y}})}{\eta(E_{\pi_J(\mathbf{y})})\xi^{\pi_J(\mathbf{x}) \cdot_s  \pi_J(\mathbf{y})}}
\ket{\psi_{\overline{J}}(\mathbf{x})}.
\]

As a consequence $\ket{\psi_{\overline{J}}(\mathbf{x})}$ is an eigenvector of
$E_{\pi_{\overline{J}}(\mathbf{y})}$ with respect to the eigenvalue
$$\frac{\lambda(E_{\mathbf{y}})}{\eta(E_{\pi_J(\mathbf{y})})\xi^{\pi_J(\mathbf{x}) \cdot_s  \pi_J(\mathbf{y})}}.$$

Thus, we have seen that $\ket{\psi_{\overline{J}}(\mathbf{x})}$
is an eigenvector of $E_{\mathbf{v}}$ for every $\mathbf{v} \in H$.
Then, for any nonvanishing $\mathbf{z} \in D_2$,
$E_{\pi_{\overline{J}}(\mathbf{z})}\ket{\psi_{\overline{J}}(\mathbf{x})}$
is also an eigenvector of $E_{\mathbf{v}}$ for every $\mathbf{v} \in H$,
and, reasoning as we did before (\ref{eq1001}), it is orthogonal to $\ket{\psi_{\overline{J}}(\mathbf{x})}$.

Next we will see that the length of
$\ket{\psi_{\overline{J}}(\mathbf{x})}$ is $1$. Let $\mathbf{z}\in D_2$, by
(\ref{eq101}) it holds
\begin{eqnarray*}
  \lambda(E_{\mathbf{z}})\ket{\varphi}
  &=& E_{\mathbf{z}}\sum_{\mathbf{x} \in D_2}\frac{1}{\sqrt{\mathrm{card}\,(D_2)}}E_{\pi_J(\mathbf{x})}\ket{\psi_J} \otimes \ket{\psi_{\overline{J}}(\mathbf{x})}\\
  &=&
  \sum_{\mathbf{x} \in D_2}\frac{1}{\sqrt{\mathrm{card}\,(D_2)}}E_{\pi_J(\mathbf{z})}E_{\pi_J(\mathbf{x})}\ket{\psi_J} \otimes E_{\pi_{\overline{J}}(\mathbf{z})}\ket{\psi_{\overline{J}}(\mathbf{x})}.
\end{eqnarray*}
Since $E_{\pi_J(\mathbf{z})}E_{\pi_J(\mathbf{x})}\ket{\psi_J}$
is a scalar (not necessarily $1$) multiple of
$E_{\pi_J(\mathbf{x})+\pi_J(\mathbf{z})}\ket{\psi_J}$,
we get that
$E_{\pi_{\overline{J}}(\mathbf{z})}\ket{\psi_{\overline{J}}(\mathbf{x})}$ is a scalar multiple of $\ket{\psi_{\overline{J}}(\mathbf{x}+\mathbf{z})}$.
Also the lengths of $\ket{\psi_{\overline{J}}(\mathbf{x})}$
and $\ket{\psi_{\overline{J}}(\mathbf{x}+\mathbf{z})}$ are the same
because $E_{\pi_{\overline{J}}(\mathbf{z})}$ is a unitary matrix.
Recall that $\ket{\psi_J}$ has length $1$.
Since vectors in the summation in (\ref{eq101})
are orthogonal to each other and their sum (\ref{eq101}) has length $1$,
it follows that, for all $\mathbf{x} \in D_2$, the length of $\ket{\psi_{\overline{J}}(\mathbf{x})}$ is $1$. Reasoning as before of (\ref{eq1001}), we also see that the vectors $\ket{\psi_{\overline{J}}(\mathbf{x})}$
are orthogonal to each other. 
We note that (\ref{eq101}) is a Schmidt decomposition \cite[Section 2.5]{chuangnielsen} of $\ket{\varphi}$.

Taking into account the previous considerations and using trace results \cite[Subsections 2.1.8 and 2.4.3]{chuangnielsen} in the forthcoming second equality, we deduce that the partial trace over $\overline{J}$ of the quantum state in (\ref{eq101})  is
\begin{eqnarray}
&&\mathrm{tr}_{\overline{J}}[\ket{\varphi}\bra{\varphi}]\nonumber\\
&=&\mathrm{tr}_{\overline{J}}\left[
\sum_{\mathbf{x}, \mathbf{x}' \in D_2}\frac{1}{\mathrm{card}\,(D_2)}E_{\pi_J(\mathbf{x})}\ket{\psi_J}\bra{\psi_J} E^\dagger_{\pi_J(\mathbf{x}')} \otimes \ket{\psi_{\overline{J}}(\mathbf{x})} \bra{\psi_{\overline{J}}(\mathbf{x}')}
\right]\nonumber\\
&=&
\sum_{\mathbf{x}, \mathbf{x}' \in D_2}\frac{1}{\mathrm{card}\,(D_2)}E_{\pi_J(\mathbf{x})}\ket{\psi_J}\bra{\psi_J} E^\dagger_{\pi_J(\mathbf{x}')} \times \underbrace{\mathrm{tr}[\ket{\psi_{\overline{J}}(\mathbf{x})} \bra{\psi_{\overline{J}}(\mathbf{x}')}]}_{=\delta_{\mathbf{x}, \mathbf{x}'}}
\nonumber\\
&=&
\sum_{\mathbf{x} \in D_2}\frac{1}{\mathrm{card}\,(D_2)}E_{\pi_J(\mathbf{x})}\ket{\psi_J}\bra{\psi_J} E^\dagger_{\pi_J(\mathbf{x})}.\label{eq1002}
\end{eqnarray}

Recalling from (\ref{eq1001}) that  $\{E_{\pi_J(\mathbf{x})}\ket{\varphi_J} \; : \; \mathbf{x} \in D_2 \}$ is
an orthonormal basis of $Q(\sigma_J(C))$ and noticing that
any $E_{\pi_J(\mathbf{x})}\ket{\psi_J}$ also belongs to that orthonormal basis because $\mathbf{x}_0$ and $\mathbf{x}$ belong to $D_2 \subseteq C$, we prove that the expression in (\ref{eq1002}) is a completely mixed state on
$Q(\sigma_J(C))$. This finishes the proof of  Lemma \ref{lemma4}.

\end{proof}

Now we state and prove the last of the aforementioned lemmas. Remind that we desire to show that Equality (\ref{NAS}) is a necessary condition for $Q(C)$ to be an $(I,J)$-locally recoverable code.

\begin{lem}
\label{elC}
Let $Q(C) \subseteq \mathbb{C}^{p^n}$ be a stabilizer code given by a symplectic self-orthogonal $p$-ary code $C \subseteq \mathbb{F}_p^{2n}$, and consider sets of indices $\emptyset \neq I \subsetneq J \varsubsetneq\{1, \ldots, n\}$. Set $\overline{J} = \{1, \ldots, n\} \setminus J$. Then, there exists an environment $\mathcal{E}$ surrounding $\mathbb{C}^{p^n}$ and an entangled quantum codeword $\ket{\Phi} \in Q(C) \otimes \mathcal{E}$
such that the partial trace of $\ket{\Phi}\bra{\Phi}$ over $\overline{J}$ and $\mathcal{E}$ is a completely mixed state on $Q(\sigma_J(C))$.
\end{lem}
\begin{proof}
Consider the code $C_{\mathrm{max}}$ defined in (\ref{max}) and its shortening $\sigma_J (C_{\mathrm{max}})$. Then $$ Q(\sigma_J (C_{\mathrm{max}})) \subseteq Q(\sigma_J(C)).$$ Furthermore, $Q(\sigma_J (C_{\mathrm{max}}))$ is orthogonal to $E_{\mathbf{z}} Q(\sigma_J (C_{\mathrm{max}}))$ if and only if $ \mathbf{z} \not \in [\sigma_J (C_{\mathrm{max}})]^{\perp_s}$.

When $\mathbf{z}$ runs over $\pi_J (C^{\perp_s}) = [\sigma_J(C)]^{\perp_s}$, the cardinality of the quotient linear space $[ (\pi_J (C^{\perp_s})/(\sigma_J (C_{\mathrm{max}}))^{\perp_s}]$ coincides with the number of different spaces $E_{\mathbf{z}} Q(\sigma_J (C_{\mathrm{max}}))$. Then, one gets the following chain of equalities:
\begin{eqnarray*}
    \dim \sum_{\mathbf{z} \in \pi_J(C^{\perp_s})} E_{\mathbf{z}} Q(\sigma_J (C_{\mathrm{max}}))
    &=& \mathrm{card}\,[ (\pi_J (C^{\perp_s})/(\sigma_J (C_{\mathrm{max}}))^{\perp_s}] \dim Q(\sigma_J(C_{\mathrm{max}}))\\
    &=& p^{\dim  (\sigma_J (C))^{\perp_s} - \dim (\sigma_J(C_{\mathrm{max}}))^{\perp_s}} \times p^{\mathrm{card}\, J - \dim \sigma_J(C_{\mathrm{max}})}\\
    &=& p^{\dim (\sigma_J(C_{\mathrm{max}}))  - \dim \sigma_J (C)} \times p^{\mathrm{card}\, J - \dim \sigma_J(C_{\mathrm{max}})} \\
    &=& p^{\mathrm{card}\, J - \dim \sigma_J (C)} = \dim Q(\sigma_J (C)).
  \end{eqnarray*}
The equality of dimensions we have just obtained and the inclusion $$\sum_{\mathbf{z} \in \pi_J(C^{\perp_s})} E_{\mathbf{z}} Q(\sigma_J (C_{\mathrm{max}})) \subseteq Q(\sigma_J (C)),$$ prove that
\begin{equation}
\label{eq301}
\sum_{\mathbf{z} \in \pi_J(C^{\perp_s})} E_{\mathbf{z}} Q(\sigma_J (C_{\mathrm{max}})) = Q(\sigma_J (C)).
\end{equation}

Recall that $(\sigma_J(C_{\mathrm{max}}))^{\perp_s} = \pi_J(C_{\mathrm{max}})$.  Consider a set $T_J \subseteq \mathbb{F}_p^{2 \mathrm{card}\,( J )}$ of representatives of the cosets in the quotient $[ \pi_J (C^{\perp_s})/\pi_J (C_{\mathrm{max}})]$.  Since  $\mathbf{z} \in (\sigma_J(C_{\mathrm{max}}))^{\perp_s}$ implies the equality of spaces  $Q(\sigma_J(C_{\mathrm{max}})) =  E_\mathbf{z} Q(\sigma_J(C_{\mathrm{max}})) $, Equality (\ref{eq301}) can be rewritten as
\begin{equation}
\label{eq302}
\bigoplus_{\mathbf{z} \in T_J } E_\mathbf{z} Q(\sigma_J (C_{\mathrm{max}})) = Q(\sigma_J (C)).
\end{equation}

Now, on the one hand, choose $T \subseteq C^{\perp_s}$ such that
$\pi_J(T) = T_J$ and $\mathrm{card}\,(T) = \mathrm{card}\,(T_J)=p^{\dim \pi_J(C^{\perp_s})- \dim \pi_J(C_{\mathrm{max}})}$. And, on the other hand, consider an environment $\mathcal{E}$ (surrounding $\mathbb{C}^{p^n}$) with an orthonormal basis $\{\ket{\mathbf{y}}_{\mathcal{E}} \; : \; \mathbf{y} \in T \}$. $T$,  $\mathcal{E}$ and any fixed  $\ket{\varphi} \in Q(C_{\mathrm{max}}) \subseteq Q(C)$  allow us to construct the announced vector $\ket{\Phi}$:
\begin{equation}
  \ket{\Phi} = \frac{1}{\sqrt{\mathrm{card}\,(T)}} \sum_{\mathbf{y}\in T}
    E_{\mathbf{y}} \ket{\varphi} \otimes \ket{\mathbf{y}}_{\mathcal{E}}.
    \label{eq303}
\end{equation}

Then, Lemma \ref{elC} follows from the subsequent sequence of equalities:
\begin{eqnarray*}
  && \mathrm{tr}_{\overline{J} \cup \mathcal{E}}\left[\ket{\Phi}\bra{\Phi}\right] \\
  &=& \mathrm{tr}_{\overline{J}}[\mathrm{tr}_{\mathcal{E}}\left[\ket{\Phi}\bra{\Phi}]
  \right]\\
  &=& \mathrm{tr}_{\overline{J}}\left[ \frac{1}{\mathrm{card}(T)}
    \sum_{\mathbf{y} \in T} E_{\mathbf{y}} \ket{\varphi}\bra{\varphi}E_{\mathbf{y}}^\dagger\right]\\
  &&
  \textrm{(because $\ket{\mathbf{y}}_{\mathcal{E}}$ is orthonormal)}\\
  &=&  \frac{1}{\mathrm{card}(T)}
  \sum_{\mathbf{y} \in T} E_{\pi_J(\mathbf{y})} \mathrm{tr}_{\overline{J}}\left[\ket{\varphi}\bra{\varphi}\right]E_{\pi_J(\mathbf{y})}^\dagger\\
  &&
  \textrm{(because the partial trace is linear)}\\
  &=&  \frac{1}{\mathrm{card}(T_J)}
  \sum_{\mathbf{z} \in T_J} E_{\mathbf{z}} \mathrm{tr}_{\overline{J}}\left[\ket{\varphi}\bra{\varphi}\right]
  E_{\mathbf{z}}^\dagger\\
  &=& \frac{1}{\mathrm{card}(T_J)}
  \sum_{\mathbf{z} \in T_J} E_{\mathbf{z}} (\textrm{a completely mixed state on }Q(\sigma_J(C_{\mathrm{max}})))E_{\mathbf{z}}^\dagger\\
  &&
  \textrm{(by Lemma \ref{lemma4})}\\
  &=& \textrm{a completely mixed state on }Q(\sigma_J(C))
  \textrm{ (by  (\ref{eq302})).}
  \end{eqnarray*}
This concludes the proof.
\end{proof}

Lemmas \ref{elC} and \ref{elA} prove our before announced result on the necessity of Condition (\ref{NAS}) for $(I,J)$-local recoverability. Let us state it.

\begin{pro}
\label{th14}
Let $Q(C) \subseteq \mathbb{C}^{p^n}$ be a stabilizer code given by a symplectic self-orthogonal $p$-ary code $C \subseteq \mathbb{F}_p^{2n}$ and consider sets of indices $\emptyset \neq I \subsetneq J \subsetneq \{1, \ldots, n\}$. The codes' equality stated in (\ref{NAS}):
\[
\sigma_I\left[\pi_J \left(C^{\perp_s}\right)\right] = \sigma_I(C)
\]
is a necessary condition for $Q(C)$ to be $(I,J)$-locally recoverable.
\end{pro}

We conclude with our main result in this subsection. It follows from Theorem \ref{twostars} and, by reasoning as in \cite[Section 2.1]{QINP3}, from Propositions \ref{th12} and \ref{th14}.

\begin{teo}
\label{IJ}
Let $Q(C) \subseteq \mathbb{C}^{q^n}$ be a stabilizer code given by a symplectic self-orthogonal $q$-ary code $C \subseteq \mathbb{F}_q^{2n}$, $q$ being a prime power. Consider sets of indices $I$ and $J$ such that $\emptyset \neq I \subsetneq J \subseteq \{1, \ldots, n\}$. Then, $Q(C)$ is $(I,J)$-locally recoverable if and only if Condition (\ref{NAS}):
\[
\sigma_I\left[\pi_J \left(C^{\perp_s}\right)\right] = \sigma_I(C)
\]
holds.
\end{teo}

\subsubsection{Generalized symplectic weight and $(I,J)$-local recoverability}
This subsection recalls the concept of generalized symplectic weight and relates it with (quantum) $(I,J)$-local recoverability. Let us start with the definition of generalized symplectic weight \cite[Definition 15]{matsumoto19qinp}.

\begin{de}\label{defgsw}
Let $C \subseteq \mathbb{F}_q^{2n}$ an $\mathbb{F}_q$-linear space. For each index $1 \leq t \leq \dim C$, the $t$-th {\it generalized symplectic weight}  of $C$, $\mathrm{gsw}_t (C)$ is defined as
\[
\mathrm{gsw}_t (C) := \min \{\mathrm{card}\, (J) : \dim (\sigma_J(C)) \geq t \}.
\]
\end{de}

Note that $\mathrm{gsw}_1 (C) = \mathrm{swt} (C)$, $\mathrm{gsw}_t (C)$ needs not be equal to $\min \{\mathrm{card}\, (J) : \dim (\sigma_J(C) = t \}$ and $\mathrm{gsw}_{t+1} (C) \geq \mathrm{gsw}_t (C)$, $1 \leq t \leq \dim C -1$.

Our first result is the following one.
\begin{pro}
\label{teo8}
Let $Q(C)$ be a stabilizer code defined by a symplectic self-orthogonal code $C \subseteq \mathbb{F}_q^{2n}$. Consider sets of indices $\emptyset \neq I \subsetneq J \subseteq \{1, \ldots, n\}$ and assume that $\mathrm{card}\, (J) \geq n - \mathrm{swt} (C^{\perp_s})
+ \mathrm{card}\, (I) +1$, then $Q(C)$ is $(I,J)$-locally recoverable.
\end{pro}
\begin{proof}
The fact that $\mathrm{swt} (C^{\perp_s}) \geq n - \mathrm{card}\, (J) + \mathrm{card}\, (I) +1$ implies that $$\mathrm{swt}[ \pi_J (C^{\perp_s})] \geq \mathrm{card}\, (I) +1,$$ and then $\sigma_I[ \pi_J (C^{\perp_s})] = \{\mathbf{0}\}$. Thus, Equality (\ref{NAS}) holds and the result follows by Theorem \ref{IJ}.
\end{proof}

Next, we give our last result in this subsection. It provides two inequalities that must hold when $Q(C)$ is $(I,J)$-locally recoverable. We state the counter-reciprocal statement.

\begin{pro}
\label{teo11}
Keep the notation as in Proposition \ref{teo8}, $Q(C)$ being an $[[n,k]]_q$ stabilizer code. Assume that $\mathrm{swt}(C) \geq \mathrm{card}\, (I) +1$. Then, Equality (\ref{NAS}) is false whenever the following two inequalities hold:
\begin{equation}
\label{was15-1}
 n + k - 2 \, \mathrm{card}\, (J) + 2 \, \mathrm{card}\, (I) > 0
\end{equation}
and
\begin{equation}
\label{was15-2}
\mathrm{gsw}_{n + k - 2 \, \mathrm{card}\, (J) + 2 \, \mathrm{card}\, (I)} (C^{\perp_s}) \geq n - \mathrm{card}\, (J) + 1.
\end{equation}
\end{pro}
\begin{proof}
The assumption $\mathrm{swt}(C) \geq \mathrm{card}\, (I) +1$ implies that the right hand side of Equality (\ref{NAS}) vanishes.

The shortening at $I$ of $\pi_J \left(C^{\perp_s}\right)$ removes $(2 \, \mathrm{card}\, (J) - 2 \, \mathrm{card}\, (I))$ components and then \begin{equation}
\label{OM}
\dim \pi_J \left(C^{\perp_s}\right) \geq 2 \, \mathrm{card}\, (J) - 2 \, \mathrm{card}\, (I) +1
\end{equation}
implies that $\sigma_I\left[\pi_J \left(C^{\perp_s}\right)\right] \neq \{\mathbf{0}\}$ and therefore Equality (\ref{NAS}) is not true.

Let us show that Inequalities (\ref{was15-1}) and (\ref{was15-2}) imply (\ref{OM}).

Consider the linear map $\pi_J: \mathbb{F}_q^{2n} \rightarrow \mathbb{F}_q^{2 \, \mathrm{card}\, (J)}$, whose kernel is $\sigma_{\overline{J}}(\mathbb{F}_q^{2n})$ and thus
\[
\dim \pi_J(C^{\perp_s}) = n+k - \dim (C^{\perp_s} \cap \sigma_{\overline{J}}(\mathbb{F}_q^{2n}))
= n+k - \dim (\sigma_{\overline{J}} (C^{\perp_s})).
\]
Then, Inequality (\ref{OM}) is equivalent to the inequality
$$
n+k - \dim (\sigma_{\overline{J}} (C^{\perp_s})) > 2 \, \mathrm{card}\, (J) - 2 \, \mathrm{card}\, (I)
$$
and this inequality is equivalent to
\begin{equation}
\label{was18}
\dim ( \sigma_{\overline{J}} (C^{\perp_s})) < n + k - 2 \, \mathrm{card}\, (J) + 2 \, \mathrm{card}\, (I).
\end{equation}

The fact that inequalities (\ref{was15-1}) and (\ref{was15-2}) imply (\ref{was18}) concludes the proof.
\end{proof}

\begin{exa}
\label{eje1}
{\rm
Contrary to what happens with classical error-correcting codes, the above inequality $\mathrm{gsw}_{t+1} (C) \geq \mathrm{gsw}_t (C)$ needs not be strict. Indeed, Let $C_H^{\perp_e}$ be the $[7,4,3]_2$ Hamming code
with $C_H \subseteq C_H^{\perp_e}$.
By \cite{wei91}, the generalized Hamming weight hierarchy of $C_H$ is $(4,6,7)$ and
that of $C_H^{\perp_e}$ is  $(3,5,6,7)$.
Consider the code $C:=C_H \times C_H \subseteq \mathbb{F}_2^{14}$,
which defines the $[[7,1,3]]_2$ Steane quantum stabilizer code.
We have $\mathrm{gsw}_1 (C^{\perp_s}) = \mathrm{gsw}_2(C^{\perp_s})= 3$ and $\mathrm{gsw}_3(C^{\perp_s}) = \mathrm{gsw}_4(C^{\perp_s}) = 5$.

In addition, Proposition \ref{teo8} proves that for any $i \in J$ with $ \mathrm{card}\, (J)=6$,
an erasure at $i$ can be recovered by $J$. 
}
\end{exa}

\subsection{Quantum $(r,\delta)$-local recoverability}
\label{section33}
This brief subsection takes the previous results to the language of quantum $(r,\delta)$-locally recoverable codes. Our results follow directly from preceding results in this section.

The first one follows straightforwardly from Theorem \ref{IJ}.

\begin{teo}
\label{quantumsyp}
Let $Q(C) \subseteq \mathbb{C}^{q^n}$ be a stabilizer code given by a symplectic self-orthogonal $q$-ary linear code $C \subseteq \mathbb{F}_q^{2n}$, $q$ being a prime power. $Q(C)$ is a quantum $(r,\delta)$-LRC if and only if for all index $i \in \{1, \ldots, n\}$, there exists a set $J \ni i$ of cardinality $\leq r + \delta -1$ such that the following equality holds
\[
\sigma_I\left[\pi_J \left(C^{\perp_s}\right)\right] = \sigma_I(C)
\]
for any subset $I \subsetneq J$ of cardinality $\delta -1$.
\end{teo}

Our second result can be deduced from Propositions \ref{teo8} and \ref{teo11}.

\begin{pro}
Let $Q(C)$ be a stabilizer code defined by a symplectic self-orthogonal linear code $C \subseteq \mathbb{F}_q^{2n}$.
\begin{enumerate}
\item Suppose that $\mathrm{swt}(C^{\perp_s}) \geq n -r +1$. Then, for $\delta \leq n+1-r$, $Q(C)$ is a quantum $(r,\delta)$-LRC.
\item Assume that $\mathrm{swt}(C) \geq \delta$. If $Q(C)$ is an $[[n,k]]_q$ quantum $(r,\delta)$-LRC such that $r+ \delta-1 \leq n$,  then either $n+k \leq 2r$ or $n+k > 2r$ and $\mathrm{gsw}_{n+k-2r} (C^{\perp_s}) < n - (r + \delta) +2$.
\end{enumerate}
\end{pro}



\section{Quantum $(r,\delta)$-LRCs from Hermitian and Euclidean inner products.} 
\label{La4}
We divide this section in three parts. The first one gives conditions for local recoverability of quantum codes defined by Hermitian or Euclidean self-orthogonal codes. In the second one, we relate classical and quantum local recoverability  while a Singleton-like bound for quantum $(r,\delta)$-LRCs coming from Hermitian and Euclidean dual-containing codes is stated in the last part.

\subsection{Quantum $(r,\delta)$-LRCs from Hermitian and Euclidean self-orthogonal codes}

As we said before, quantum stabilizer codes can be constructed by using Hermitian or Euclidean inner products instead of symplectic forms. In these cases, Theorems \ref{IJ} and \ref{quantumsyp} can be straightforwardly translated.

Recall that the {\it Hermitian product} of two vectors $\mathbf{x}, \mathbf{y}$ in $\mathbb{F}_{q^2}^n$ is defined as $\mathbf{x} \cdot_h \mathbf{y} = \mathbf{x}^q \cdot_e \mathbf{y}$, where $\mathbf{x}^q$ is the componentwise $q$-th power of $\mathbf{x}$ and  $\cdot_e$ stands for the Euclidean inner product. Also, take a normal basis $\{\varsigma, \varsigma^q\}$ of $\mathbb{F}_{q^2}$ over $\mathbb{F}_{q}$ and remember that the {\it trace-alternating product} of the previous vectors $\mathbf{x}, \mathbf{y}$ is defined as $\mathbf{x} \cdot_a \mathbf{y} := \mathrm{tr}_{q/p} \left( \frac{ \mathbf{x}\cdot_e \mathbf{y}^q - \mathbf{x}^q \cdot_e \mathbf{y}}{\varsigma^{2q}-\varsigma^2 }\right)$. Consider the one to one isometry $\psi: \mathbb{F}_{q}^{2n} \rightarrow \mathbb{F}_{q^2}^n$ defined by $\psi(\mathbf{a}| \mathbf{b}) = \varsigma \mathbf{a} + \varsigma^q \mathbf{b}$ and then
\[
(\mathbf{a}| \mathbf{b}) ._{\mathrm{trs}} (\mathbf{c}| \mathbf{d}) = \psi\left((\mathbf{a}| \mathbf{b})\right) \cdot_a \psi \left((\mathbf{c}| \mathbf{d})\right).
\]

By \cite[Lemma 18]{Ketkar}, an $\mathbb{F}_{q^2}$-linear subspace $D$ of $\mathbb{F}_{q^2}^n$ satisfies $D^{\perp_a} = D^{\perp_h}$ and, by \cite[Theorem 15]{Ketkar}, the existence of a quantum stabilizer code is equivalent to that of an additive code $D$ in $\mathbb{F}_{q^2}^n$ satisfying $D \subseteq D^{\perp_a}$. These results allow us to translate ours on local recoverability to stabilizer codes defined by self-orthogonal, with respect to the Hermitian inner product, linear subspaces of $\mathbb{F}_{q^2}^n$.

The same happens when considering  $\mathbb{F}_{q}$-linear subspaces $C_1$ and $C_2$ included in $\mathbb{F}_{q}^n$ such that $C_2 \subseteq C_1^{\perp_e}$. Note that, in this case, one should consider $C_1 \times C_2$ and, by the CSS procedure, use the symplectic construction. In Theorem \ref{IJ-HE} we will use $C_1=C_2$ as $C$ for the Euclidean inner product.

Let us state our results.

\begin{teo}
\label{IJ-HE}
Let $Q(C) \subseteq \mathbb{C}^{q^n}$ be a stabilizer code given by a $q^2$-ary (respectively, $q$-ary) Hermitian (respectively, Euclidean) self-orthogonal linear code $C \subseteq \mathbb{F}_{q^2}^n$ (respectively, $C \subseteq \mathbb{F}_{q}^n$), $q$ being a prime power. Consider sets of indices $I$ and $J$ such that $\emptyset \neq I \subsetneq J \subseteq \{1, \ldots, n\}$. Then, $Q(C)$ is $(I,J)$-locally recoverable if and only if the equality
\begin{equation}
\label{NAS-H}
\sigma_I\left[\pi_J \left(C^{\perp_h}\right)\right] = \sigma_I(C)
\end{equation}
(respectively,
\begin{equation}
\label{NAS-E}
\sigma_I\left[\pi_J \left(C^{\perp_e}\right)\right] = \sigma_I(C)\,)
\end{equation}
holds.
\end{teo}

\begin{teo}
\label{quantumH-E}
Let $Q(C) \subseteq \mathbb{C}^{q^n}$ be a stabilizer code given by a $q^2$-ary (respectively $q$-ary) Hermitian (respectively, Euclidean)  self-orthogonal code $C \subseteq \mathbb{F}_{q^2}^n$ (respectively, $C \subseteq \mathbb{F}_{q}^n$), $q$ being a prime power. $Q(C)$ is a quantum $(r,\delta)$-LRC if and only if, for all index $i \in \{1, \ldots, n\}$, there exists a set $J \ni i$ of cardinality $\leq r + \delta -1$ such that  Equality (\ref{NAS-H}) (respectively,  Equality (\ref{NAS-E})) holds for any subset $I \subsetneq J$ of cardinality $\delta -1$.
\end{teo}

Theorem \ref{IJ-HE} contemplates a unique code $C$ in the CSS case. For the general CSS construction we state the following result.

\begin{pro}
Let $Q(C) \subseteq \mathbb{C}^{q^n}$ be a stabilizer code given by two codes $C_1$ and $C_2$ included in $\mathbb{F}_{q}^n$ and such that $C_2 \subseteq C_1^{\perp_e}$. Then, $Q(C)$ is $(I,J)$-locally recoverable if and only if the following equalities
\[
\sigma_I\left[\pi_J \left(C_1^{\perp_e}\right)\right] = \sigma_I(C_2)
\]
and
\[
\sigma_I\left[\pi_J \left(C_2^{\perp_e}\right)\right] = \sigma_I(C_1)
\]
hold.
\end{pro}
\begin{proof}
By  Theorem \ref{IJ}, $Q(C)$ is $(I,J)$-locally recoverable if and only if
\[
\sigma_I\left[\pi_J \left(\left(C_1 \times C_2\right)^{\perp_s}\right)\right]  = \sigma_I[(C_1 \times C_2)]
\]
which is equivalent to
\[
\sigma_I\left[\pi_J \left(C_2^{\perp_e}\right)\right]  \times \sigma_I\left[\pi_J \left(C_1^{\perp_e}\right)\right] = \sigma_I(C_1) \times \sigma_I(C_2),
\]
which is equivalent to the conditions in the statement. This ends the proof.
\end{proof}

\subsection{Quantum and classical $(r,\delta)$-LRCs}
\label{subsect42}

We start by stating and proving the following result concerning $(I,J)$-local recoverability of classical codes.

\begin{pro}
\label{classical}
Let $C$ be a $q^2$-ary linear code included in $\mathbb{F}_{q^2}^n$ (respectively, a $q$-ary linear code included in $\mathbb{F}_{q}^n$). Consider sets of indices $I$ and $J$ such that $\emptyset \neq I \subsetneq J \subseteq \{1, \ldots, n\}$. Then, erasures of the code $C$ at $I$ can be classically corrected from the knowledge of the coordinates at $J \setminus I$ if and only if the following equivalent conditions hold:
\begin{equation}
\label{classical-H}
\sigma_I \left[ \pi_J(C) \right] = \{\mathbf{0}\} \Longleftrightarrow \sigma_I \left[ \left(\sigma_J(C^{\perp_h})\right)^{\perp_h} \right] = \{\mathbf{0}\}
\end{equation}
(respectively,
\begin{equation}
\label{classical-E}
\sigma_I \left[ \pi_J(C) \right] = \{\mathbf{0}\} \Longleftrightarrow \sigma_I \left[ \left(\sigma_J(C^{\perp_e})\right)^{\perp_e} \right] = \{\mathbf{0}\} \,).
\end{equation}
\end{pro}

\begin{proof}
We give a proof for the Euclidean case; the Hermitian case follows analogously. Suppose  that $C$ is an $[n,k]_q$ code. Take a basis $\{\mathbf{h}_1, \ldots, \mathbf{h}_{n-k}\}$ of the dual space $C^{\perp_e}$ and consider the map $f: \mathbb{F}_{q}^n \rightarrow \mathbb{F}_{q}^{n-k}$ defined as $f(\mathbf{x}) =(\mathbf{x} \cdot_e \mathbf{h}_1, \ldots, \mathbf{x} \cdot_e \mathbf{h}_{n-k})$. Correction of erasures consists of solving a system of linear equations $f(\mathbf{e}) = \mathbf{s}$ where $\mathbf{e} = (e_1, \ldots, e_n)$ is unknown and $\mathbf{s} = (s_1, \ldots, s_{n-k})$ a syndrome obtained from the received word. Since erasures are in $I$, we have $e_i = 0$ whenever $i \notin I$ and noticing that $\ker (f) = C$, we get a unique solution if and only if $\sigma_I(C) = \{\mathbf{0}\}$. Finally, we only use coordinates at $J$ for local recovery, which means that we use $\pi_J (C)$ instead of $C$ and the result is proved.
\end{proof}

Our next result relates $(I,J)$-local recoverability of quantum and classical codes. We assume that the classical codes satisfy dual-containing instead of self-orthogonality because, under this assumption, distances of quantum codes fit better with those of classical codes. Recall that, as we said in the introduction, we denote by $Q'(C)$ the quantum  stabilizer code given by a Hermitian or Euclidean dual-containing code $C$. Note that $Q'(C) = Q(C^\perp)$, where $C^\perp$ is the corresponding Hermitian or Euclidean dual code of $C$.

\begin{pro}
\label{clas-quant}
Let $C$ be a $q^2$-ary linear code included in $\mathbb{F}_{q^2}^n$ (respectively, a $q$-ary linear code included in $\mathbb{F}_{q}^n$). Assume that $C^{\perp_h} \subseteq C$ (respectively, $C^{\perp_e} \subseteq C$) and that $\dim C = \frac{n+k}{2}$. Consider sets of indices $I$ and $J$ such that $\emptyset \neq I \subsetneq J \subseteq \{1, \ldots, n\}$. Assume that $\mathrm{card}\, (I) \leq d(C^{\perp_h} ) -1$ (respectively, $\mathrm{card}\, (I) \leq d(C^{\perp_e} ) -1$).

Then, $Q'(C)$ is an $(I,J)$-locally recoverable code if and only if erasures of the code $C$ at $I$ can be classically corrected from the knowledge of the coordinates of $C$ at $J \setminus I$.
\end{pro}

\begin{proof}
The assumption $\mathrm{card}\, (I) \leq d(C^{\perp_h} ) -1$ (respectively, $\mathrm{card}\, (I) \leq d(C^{\perp_e} ) -1$) implies
$\sigma_I \left(C^{\perp_h} \right) = \{\mathbf{0}\}$ (respectively, $\sigma_I \left( C^{\perp_e} \right) = \{\mathbf{0}\}$. Then, noticing that we are considering dual-containing and the projection $\pi_J$, Equality (\ref{classical-H})  (respectively, Equality (\ref{classical-E})) is equivalent to Equality (\ref{NAS-H}) (respectively, Equality (\ref{NAS-E})), which concludes the proof.
\end{proof}

We finish this subsection with our main result relating quantum and classical $(r,\delta)$-locally recoverable codes. It follows from Proposition \ref{clas-quant}.

\begin{teo}
\label{relation}
Let $C$ be a $q^2$-ary  linear code included in $\mathbb{F}_{q^2}^n$ (respectively, a  $q$-ary linear code included in $\mathbb{F}_{q}^n$). Assume that $C$ is Hermitian  (respectively, Euclidean) dual-containing. Suppose also that $\delta \leq d(C^{\perp_h})$ (respectively, $d(C^{\perp_e}))$. Then, $Q'(C)$ is a quantum  $(r,\delta)$-LRC if and only if $C$ is a classical  $(r,\delta)$-LRC.
\end{teo}

\begin{rem}
Corollary 3 in  \cite{LuoG} proves that a  Euclidean dual-containing classical LRC of locality $r$ provides a quantum LRC of locality $r$.
\end{rem}

\subsection{A Singleton-like bound}
Our last subsection provides a Singleton-like bound for quantum $(r,\delta)$-LRCs coming from Hermitian or Euclidean dual-containing classical codes. 

\begin{teo}
\label{SingletonQ}
Let $C$ be a $q^2$-ary  linear code included in $\mathbb{F}_{q^2}^n$ (respectively, a  $q$-ary linear code included in $\mathbb{F}_{q}^n$). Assume that $C$ is Hermitian  (respectively, Euclidean) dual-containing, $\dim C = \frac{n+k}{2}$ and  $C$ is an $(r,\delta)$-LRC. Then, the quantum $(r,\delta)$-LRC, $Q'(C)$, has parameters $[[n,k, \geq d(C)]]_q$, which satisfy
\begin{equation}
\label{eq17}
k + 2 d(C) + 2\left(\left\lceil\frac{n+k}{2r}\right\rceil-1\right)(\delta-1) \leq n+2.
\end{equation}
\end{teo}

\begin{proof}
The fact that $C$ is a classical $(r,\delta)$-LRC and also  Hermitian  (respectively, Euclidean) dual-containing shows that $\delta \leq d(C) \leq d(C^{\perp_h})$ (respectively, $\delta \leq d(C) \leq d(C^{\perp_e})$). Then, Theorem \ref{relation} proves that $Q'(C)$ is a quantum  $(r,\delta)$-LRC. Using again that $C$ is a classical $(r,\delta)$-LRC, one deduces that their  parameters satisfy Inequality (\ref{Singleform}). Taking into account that $\dim C = \frac{n+k}{2}$, we get
the inequality
\[
\frac{n+k}{2}+d(C) + \left(\left\lceil\frac{n+k}{2r}\right\rceil-1\right)(\delta-1) \leq n+1,
\]
which is equivalent to (\ref{eq17}) and, thus, the result is proved.
\end{proof}

\begin{rem}
Inequality (\ref{eq17}) also holds if the assumption that $C$ is an $(r,\delta)$-LRC in Theorem \ref{SingletonQ} is replaced by $Q'(C)$ being a quantum  $(r,\delta)$-LRC and $\delta \leq d(C^{\perp_h})$ (respectively, $d(C^{\perp_e}))$.
\end{rem}

\begin{de}
{\rm A pure quantum $(r, \delta)$-LRC  $Q'(C)$ is said to be {\it optimal} whenever its parameters and $(r, \delta)$-locality attain the bound in (\ref{eq17}).}
\end{de}

\begin{rem}
Inequality (\ref{eq17}) depends on the minimum distance of the code $C$ and not on the minimum distance of $Q'(C)$. When considering pure codes $Q'(C)$ one can replace $d(C)$ with $d(Q'(C))$ in (\ref{eq17}), however in the non-pure case, setting  $d(Q'(C))$ instead of $d(C)$, the bound $n+2$ could be exceeded. We do not know a value to replace $n+2$ in this case, which would allow us to establish a general concept of optimality of quantum $(r, \delta)$-LRCs.

\end{rem}

\section{Examples}
\label{examples}

We divide this section in two subsections. The first one gives some results on and examples of optimal Euclidean dual-containing (classical) $(r,\delta)$-LRCs, $C$, giving rise to  optimal pure quantum
$(r,\delta)$-LRCs, $Q'(C)$. Our second subsection does the same but in the case of considering Hermitian duality.

\subsection{Some examples of optimal pure quantum $(r,\delta)$-LRCs by using Euclidean duality}
\label{51}
The proof of Theorem \ref{SingletonQ} shows that it suffices to provide an optimal Euclidean dual-containing $(r,\delta)$-LRC, $C$, giving rise to a pure quantum stabilizer code, for having an optimal (pure) quantum $(r,\delta)$-LRC, $Q'(C)$, and determining  its parameters.

Taking into account this fact, in this subsection we are going to give some families of $\emptyset$-affine variety codes giving rise to suitable optimal $(r,\delta)$-LRCs and therefore to  optimal pure quantum $(r,\delta)$-LRCs. $\emptyset$-affine variety codes were introduced in \cite{QINP2} and they are monomial-Cartesian codes as indicated in \cite{GFMC}. Let us recall the definition of these codes. 

Remind that $q$ is a power of a prime number $p$ and consider two integer numbers $n_1$ and $n_2$ such that both $n_1-1$ and  $n_2-1$ divide $q-1$. Let $\mathcal{J}$ be the ideal of the polynomial ring $\mathbb{F}_q[X,Y]$ generated by the binomials $X^{n_1}- X$ and $Y^{n_2}-Y$.  Set
$$P = \{\boldsymbol{\alpha}_1,\dots,\boldsymbol{\alpha}_n\} \subseteq \mathbb{F}_q^2$$
the zero-set of $\mathcal{J}$.

Let
$$E=\{0,1,\dots,n_1-1\}\times\{0,1,\dots,n_2-1\}$$
be the set of pairs corresponding to exponents of the monomials whose classes we are going to evaluate with the following linear evaluation map:
$$
\mathrm{ev}_P: \frac{\mathbb{F}_q[X,Y]}{\mathcal{J}} \rightarrow \mathbb{F}_q^{n} \textrm{, } \quad \mathrm{ev}_P(f)=\left(f(\boldsymbol{\alpha}_1),\dots,f(\boldsymbol{\alpha}_n)\right).
$$
Then, for each set $ \emptyset \neq \Delta \subseteq E$, we define the $\emptyset$-affine variety code
$$C_\Delta^P:=\langle \mathrm{ev}_P(X^{e_1}Y^{e_2}) \;: \; (e_1,\,e_2)\in \Delta\rangle \subseteq \mathbb{F}_q^n,$$
where $ \langle W \rangle$ means the $\mathbb{F}_q$-linear space generated by the elements in $W \subseteq \mathbb{F}_q^n$.

Now we are ready to show our results.

\begin{figure}[h]
    \centering
    \begin{subfigure}[b]{0.45\textwidth}
        \centering
        \begin{tikzpicture}[y=0.7cm, x=0.7cm,font=\normalsize]

        \filldraw[fill=gray!30] (0,0) rectangle (2,4);
        \draw (2,0) -- (4,0);

        \node [below] at (0,0) {\scriptsize$0$};
        \node [below] at (1,0) {$\dots$};
        \node [below] at (2,0) {\scriptsize$i$};
        \node [below] at (3,0) {$\dots$};
        \node [below] at (4,0) {\scriptsize$n_1-1$};
        \node [left] at (0,0) {\scriptsize$0$};
        \node [left] at (0,1) {$\vdots$};
        \node [left] at (0,2) {\scriptsize$j$};
        \node [left] at (0,3) {$\vdots$};
        \node [left] at (0,4) {\scriptsize$n_2-1$};

        \end{tikzpicture}
        \caption{Sets $\Delta_{i,n_2-1}$}
    \end{subfigure}
    \hfill
    \begin{subfigure}[b]{0.45\textwidth}
        \centering
        \begin{tikzpicture}[y=0.7cm, x=0.7cm,font=\normalsize]

        \filldraw[fill=gray!30] (0,0) rectangle (4,2);
        \draw (0,2) -- (0,4);

        \node [below] at (0,0) {\scriptsize$0$};
        \node [below] at (1,0) {$\dots$};
        \node [below] at (2,0) {\scriptsize$i$};
        \node [below] at (3,0) {$\dots$};
        \node [below] at (4,0) {\scriptsize$n_1-1$};
        \node [left] at (0,0) {\scriptsize$0$};
        \node [left] at (0,1) {$\vdots$};
        \node [left] at (0,2) {\scriptsize$j$};
        \node [left] at (0,3) {$\vdots$};
        \node [left] at (0,4) {\scriptsize$n_2-1$};

        \end{tikzpicture}
        \caption{Sets $\Delta_{n_1-1,j}$}
    \end{subfigure}
    \caption{Sets $\Delta_{i,n_2-1}$ and $\Delta_{n_1-1,j}$ in Proposition \ref{rect}}
    \label{fig:rect}
\end{figure}
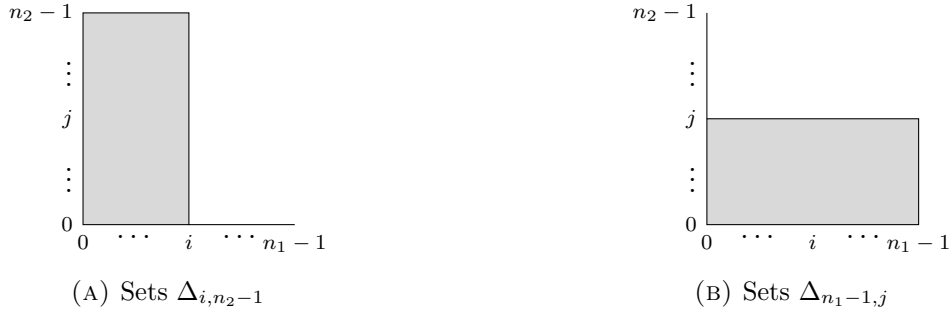

\begin{pro}\label{rect}
Keep the above notation and assume that $p$ divides $n_1$ and $n_2$. For each pair $(i,j) \in E$, consider the set
$$\Delta=\Delta_{i,j}:=\left\{(e_1,e_2) \mid 0\leq e_1 \leq i \textrm{, } 0\leq e_2\leq j\right\}\subseteq E=\{0,\dots,n_1-1\}\times\{0,\dots,n_2-1\}.$$

Then, the  code $C_\Delta^P$ is a Euclidean dual-containing  optimal $(r,\delta)$-LRC
and gives rise to  an optimal pure quantum $(r,\delta)$-LRC, $Q'(C_\Delta^P)$, if one of the following conditions hold:
\begin{itemize}
\item $i > \frac{n_1}{2}$ and $j=n_2-1$, in which case $(r,\delta)=(i+1,n_1-i)$ (see Figure \ref{fig:rect} (A)).

\item $i=n_1-1$ and $j >\frac{n_2}{2}$, in which case $(r,\delta)=(j+1,n_2-j)$ (see Figure \ref{fig:rect} (B)).
\end{itemize}

As a consequence, for any tuple $(n_1,n_2,i,j)$ of values as above described, there exists an optimal pure quantum $(r,\delta)$-LRC, $Q'(C_\Delta^P)$, with parameters
$$[[n_1n_2, 2(i+1)(j+1)- n_1n_2, (n_1-i)(n_2-j)]]_q$$
and localities as above indicated.
\end{pro}
\begin{proof}
Let us prove the first item. The second one holds by symmetry.

First, recall that the length of $C_\Delta^P$ coincides with the cardinality of $E$, its dimension is equal to the cardinality of $\Delta$ and the minimum distance can be deduced from the footprint bound and the fact that $\Delta$ is a decreasing set in the sense of \cite[Section 3]{GFMC}.
The fact that $C_\Delta^P$ is an optimal $(r,\delta)$-LRC follows from \cite[Proposition 4.1]{GFMC}. Now, \cite[Proposition 2.2]{QINP2} proves that, under the conditions in the statement, $(C_\Delta^P)^{\perp_e} \subsetneq C_\Delta^P$. Notice that $(C_\Delta^P)^{\perp_e} = C_{\Delta_{n_1-i,j}}^P$ and $\delta =d(C_\Delta^P)=n_1-i <i=d\left((C_\Delta^P)^{\perp_e}\right)$, which proves that the derived quantum code $Q'(C_\Delta^P)$ is pure.
Then, all the requirements in Theorem \ref{relation} are satisfied and $Q'(C_\Delta^P)$ is an optimal pure quantum $(r,\delta)$-LRC. The parameters of $Q'(C_\Delta^P)$ follow from Item (2) in Theorem \ref{resto}.
\end{proof}

Our second result can be proved with the same arguments we used to prove Proposition \ref{rect}, but, instead of Proposition 4.1,  we need to use Proposition 4.2 in \cite{GFMC}.

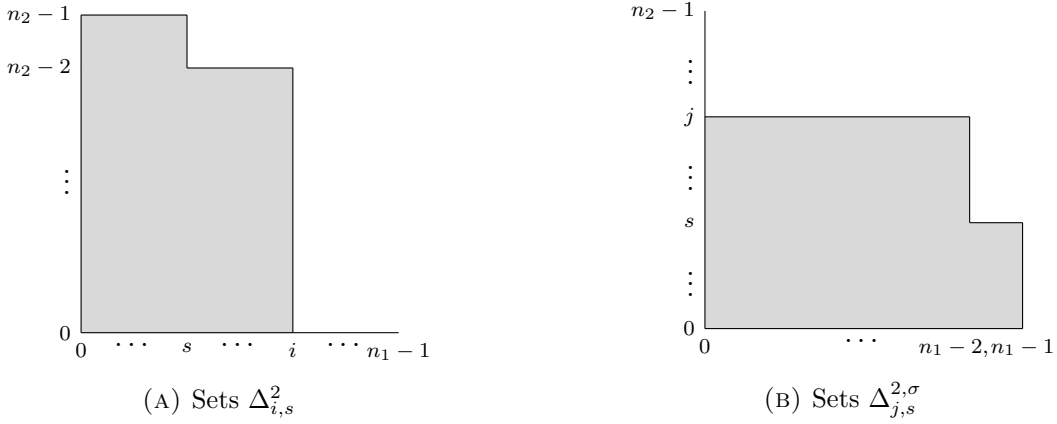
\begin{figure}[h]
    \centering
    \begin{subfigure}[b]{0.45\textwidth}
        \centering
        \begin{tikzpicture}[y=0.7cm, x=0.7cm,font=\normalsize]

        \filldraw[gray!30] (0,0) rectangle (4,5);
        \filldraw[gray!30] (0,5) rectangle (2,6);
        \draw (0,0) -- (6,0);
        \draw (0,0) -- (0,6);
        \draw (0,6) -- (2,6);
        \draw (2,6) -- (2,5);
        \draw (2,5) -- (4,5);
        \draw (4,0) -- (4,5);


        \node [below] at (0,0) {\scriptsize$0$};
        \node [below] at (1,0) {$\dots$};
        \node [below] at (2,0) {\scriptsize$s$};
        \node [below] at (3,0) {$\dots$};
        \node [below] at (4,0) {\scriptsize$i$};
        \node [below] at (5,0) {$\dots$};
        \node [below] at (6,0) {\scriptsize$n_1-1$};
        \node [left] at (0,0) {\scriptsize$0$};
        \node [left] at (0,3) {$\vdots$};
        \node [left] at (0,5) {\scriptsize$n_2-2$};
        \node [left] at (0,6) {\scriptsize$n_2-1$};

        \end{tikzpicture}
        \caption{Sets $\Delta_{i,s}^2$}
    \end{subfigure}
    \hfill
    \begin{subfigure}[b]{0.45\textwidth}
        \centering
        \begin{tikzpicture}[y=0.7cm, x=0.7cm,font=\normalsize]

        \filldraw[gray!30] (0,0) rectangle (5,4);
        \filldraw[gray!30] (5,0) rectangle (6,2);
        \draw (0,0) -- (0,6);
        \draw (0,0) -- (6,0);
        \draw (6,0) -- (6,2);
        \draw (6,2) -- (5,2);
        \draw (5,2) -- (5,4);
        \draw (0,4) -- (5,4);


        \node [left] at (0,0) {\scriptsize$0$};
        \node [left] at (0,1) {$\vdots$};
        \node [left] at (0,2) {\scriptsize$s$};
        \node [left] at (0,3) {$\vdots$};
        \node [left] at (0,4) {\scriptsize$j$};
        \node [left] at (0,5) {$\vdots$};
        \node [left] at (0,6) {\scriptsize$n_2-1$};
        \node [below] at (0,0) {\scriptsize$0$};
        \node [below] at (3,0) {$\dots$};
        \node [below] at (4.7,0) {\scriptsize$n_1-2,$};
        \node [below] at (6,0) {\scriptsize$n_1-1$};

        \end{tikzpicture}
        \caption{Sets $\Delta_{j,s}^{2,\sigma}$}
    \end{subfigure}
    \caption{Sets $\Delta_{i,s}^2$ and $\Delta_{j,s}^{2,\sigma}$ in Proposition \ref{rectelim}}
    \label{fig:rectelim}
\end{figure}

\begin{pro}\label{rectelim}
Keep the same notation and assumption as in Proposition \ref{rect}. Consider the set
$$\Delta=\Delta_{i,s}^2:=\left\{(e_1,e_2) \; : \; 0\leq e_1 \leq i
 \textrm{, } 0\leq e_2 \leq n_2-2\right\}\cup \left\{(e_1,n_2-1) \mid 0\leq e_1\leq s\right\},$$
where $\frac{n_1-1}{2}< i\leq n_1-2$ and $i > s \geq \max\{n_1-i-1,2i-n_1\}$ (see Figure \ref{fig:rectelim} (A)).

Then, the  code $C_{\Delta}^P$
gives rise to  an optimal pure quantum $(r,\delta)=(i+1,n_1-i)$-LRC, $Q'(C_\Delta^P)$.

As a consequence, for any tuple $(n_1,n_2,i,s)$ of values as above described, there exists an optimal pure quantum $(r,\delta)$-LRC, $Q'(C_\Delta^P)$, with parameters
$$[[n_1n_2, 2[(i+1)(n_2-1)+s+1]- n_1n_2, (n_1-s)]]_q$$
and the above mentioned locality.\\

Analogously, the  code $C_{\Delta}^P$, where
$$\Delta=\Delta_{j,s}^{2,\sigma}:=\left\{(e_1,e_2) \; : \; 0\leq e_1 \leq n_1-2 \textrm{, } 0\leq e_2 \leq j
\right\}\cup \left\{(n_1-1,e_2) \mid 0\leq e_2\leq s\right\},$$
for $\frac{n_2-1}{2}< j\leq n_2-2$ and $j > s \geq \max\{n_2-j -1, 2j-n_2\}$ (see Figure \ref{fig:rectelim} (B))
gives rise to  an optimal pure quantum $(r,\delta)=(j+1,n_2-j)$-LRC, $Q'(C_{\Delta}^P)$.

As a consequence, for any tuple $(n_1,n_2,j,s)$ of values as above described, there exists an optimal pure quantum $(r,\delta)$-LRC, $Q'(C_\Delta^P)$, with parameters
 $$[[n_1n_2, 2[(j+1)(n_1-1)+s+1]- n_1n_2, (n_2-s)]]_q$$
 and the before mentioned locality.
\end{pro}

\begin{exa}
Consider the values $q=7$, $n_1=n_2=7$, $i=5$ and $j=6$. Set $\Delta = \Delta_{5,6}$ as defined in Proposition \ref{rect}. Since $i=5 > \frac{7}{2}$, Proposition \ref{rect} proves that $Q'(C_\Delta^P)$ is an optimal pure quantum $(r, \delta)= (6,2)$-LRC with parameters $[[49,35, 2]]_7$ and therefore their parameters and locality reach the bound in (\ref{eq17}). Notice that $Q'(C_\Delta^P)$  is also a quantum LRC with locality $6$ attaining the bound (\ref{SingletonRQ}).
\end{exa}

\subsection{Some examples of optimal pure quantum $(r,\delta)$-LRCs by using Hermitian duality}
\label{52}
We are going to show that $(r, \delta)$-LRCs coming from Hermitian dual-containing MDS $q^2$-ary classical codes provide optimal pure quantum $(r, \delta)$-LRCs. Note that the same happens when considering MDS $q$-ary  classical codes and Euclidean dual containment.


\begin{pro}
\label{Z}
Let $C$ be an $[n,k,n-k+1]_{q^2}$ Hermitian dual-containing classical code. Then, $C$ is an optimal $(r, \delta) = (k, n-k+1)$-LRC giving rise to a pure quantum code $Q'(C)$.
Therefore, 
$Q'(C)$ is an optimal pure quantum $(r, \delta) = (k, n-k+1)$-LRC with parameters $[[n, 2k-n, n-k+1]]_q$.
\end{pro}
\begin{proof}
$C$ is an MDS code and thus $C^{\perp_h}$ has parameters $[n, n-k, k+1]_{q^2}$. In addition, $C$ is an $(r, \delta) = (k, n-k+1)$-LRC because $r + \delta -1 =n$ and the parity-check matrix of $C$ has size $(n-k) \times n$ and maximum rank, which shows that it allows to recover $\delta -1$ erasures. 
Now, $C$ is a $(k,n-k+1)$-LRC and thus $\delta \leq d(C^{\perp_h})$.
The fact that $Q'(C)$ is an optimal pure quantum $(r, \delta) = (k, n-k+1)$-LRC follows from Theorems \ref{relation} and \ref{SingletonQ}. 
This concludes the proof.
\end{proof}

\begin{rem}
{\rm Proposition \ref{Z} (and its obvious analogue for MDS $q$-ary classical codes) show that any MDS   quantum stabilizer code, obtained from  a classical code satisfying Hermitian (or Euclidean) dual containment, is optimal pure for the locality described in the statement. This is because these stabilizer codes always come from a dual-containing code where $2k -n \geq 0$. Specific parameters of many MDS stabilizer codes can be consulted in \cite{wang} and references therein.}
\end{rem}




\section*{Conflict of interest}
The authors declare they have no conflict of interest.

\section*{Acknowledgements}
The authors thank the anonymous reviewers for their helpful comments and feedback, which  improved this manuscript. We are particularly indebted with one of the reviewers. He/she found a mistake in a previous proof of Lemma 15 and helped us in defining optimality of pure codes.


\end{document}